\documentclass[12pt]{article}

\usepackage{amsmath,graphicx}
\usepackage{epstopdf} 
\usepackage{multirow}
\usepackage{amsfonts}
\usepackage{amssymb}
\usepackage{amscd}

\def\hybrid{\topmargin -20pt    \oddsidemargin 0pt
        \headheight 0pt \headsep 0pt
        \textwidth 6.25in       
        \textheight 9.5in       
        \marginparwidth .875in
        \parskip 5pt plus 1pt   \jot = 1.5ex}

\hybrid
\usepackage{xy}
\input xy 
\xyoption{all}
\usepackage{amsmath}
\usepackage{hyperref}
\usepackage{graphicx}
\usepackage{color}
\usepackage{amssymb}
\usepackage{esint}
\usepackage{amsmath, amsthm, amssymb}
\usepackage{amsmath}
\usepackage{setspace}
\numberwithin{equation}{section}
\makeatletter

\newcommand{\sla}{\slash\!\!\!}
\newcommand{\cl}{\mathcal}

\def\beq{\begin{equation}}
\def\eeq{\end{equation}}
\def\beqa{\begin{eqnarray}}
\def\eeqa{\end{eqnarray}}

\def\Tr{{\rm Tr \,}}

\def\cd{{\cal D}}

\def\cj{{\cal J}}

\def\cf{{\cal F}}

\def\mg{m_{3/2}}
\def\mg2{m^2_{3/2}}

\def\Dsl{\,\raise.15ex\hbox{/}\mkern-13.5mu D} 
\def\rep#1{\mbox{{\bf #1}}}

\def\Phid{\Phi_D}
\def\cA{A}

\def\Deltam{\Delta_m}
\def\Deltan{\Delta_n}
\def\Deltap{\Delta_{p}}

\def\Deltamu{\Delta_e}
\def\Deltaphi{\Delta_d}

\newcommand{\bbR}{\mathbb{R}}

\DeclareMathOperator{\SL}{\mathit{SL}}

\DeclareMathOperator{\E7}{\mathit{E}_{7}}
\DeclareMathOperator{\Es7}{\mathit{E}_{7(7)}}

\newcommand{\Tsub}{O(6,6) \times \SL(2,\bbR)}

\newcommand{\SLR}{\SL(2,\bbR)}
\newcommand{\SLE}{\SL(8,\bbR)}
\DeclareMathOperator{\Ex6}{\mathit{E}_{6(2)}}
\newcommand{\mukai}[2]{\big<{#1},{#2}\big>}

\newcommand{\nn}{\nonumber}

\newcommand{\FI}{F_{\rm i}}
\newcommand{\FE}{F_{\rm e}}
\newcommand{\FD}{F_{\rm d}}
\newcommand{\PE}{\partial_{\rm e}}

\begin{document}

\begin{titlepage}
\begin{center}

\rightline{\small IPhT-T11/127}
\vskip 1.4cm

{\Large \bf  $\mathcal{N}=1$ vacua in Exceptional Generalized Geometry}

\vskip 1.2cm

{\bf Mariana Gra{\~n}a, Francesco Orsi }

\vskip 0.4cm

{\em Institut de Physique Th\'eorique,                   
CEA/ Saclay \\
91191 Gif-sur-Yvette Cedex, France}  \\
\vskip 0.2cm
{\tt mariana.grana@cea.fr, francesco.orsi@cea.fr}

\vskip 0.4cm
\vspace{15mm}
\begin{center} {\bf ABSTRACT } \end{center}
\vspace{2mm}
\end{center}
We study $\mathcal{N}=1$ Minkowski vacua in compactifications of type II string theory in the language of exceptional generalized geometry (EGG).  We find the differential equations governing the  EGG analogues of the pure spinors of generalized complex geometry, namely the structures which parameterize the vector and hypermultiplet moduli spaces of
the effective four-dimensional $\mathcal{N}=2$ supergravity obtained after compactification. In order to do so,  we identify a twisted differential operator that contains NS and RR fluxes and transforms covariantly under the $U$-duality group, $E_{7(7)}$. We show that the conditions for $\mathcal{N}=1$ vacua correspond to a subset of the structures being closed under the twisted derivative. 
\vfill

\today

\end{titlepage}

{\small
\tableofcontents{}
}

\begin{section}{Introduction}
\label{Intro}

Since the seminal paper of Candelas, Horowitz, Strominger and Witten \cite{CHSW}, the geometrical perspective in compactifications of string theory from ten to four dimensions had great insights. Supersymmetry conditions have been shown to constrain the  allowed internal manifolds to certain specific classes. When there are no fluxes, the internal spaces should be Calabi-Yau. Such manifolds satisfy an  \textit{algebraic} condition, namely the existence of a globally defined, nowhere vanishing, internal spinor, and a \textit{differential} one, that the spinor is covariantly constant. The algebraic condition is necessary in order to recover a supersymmetric ($\mathcal{N}=2$) effective \textit{theory} in four dimensions, while the differential one is required in order to have supersymmetric \textit{vacua}.  In the presence of fluxes, the algebraic condition stays intact (i.e., in order to have $\mathcal{N}=2$ supersymmetry off-shell, a globally defined internal spinor is needed), but the differential one becomes more intricate.  

The role of fluxes in string theory, combined with the warped nature of the compactification, has become of primary interest mainly for the possibility of fixing moduli and providing a hierarchy of scales \cite{GKP}. This motivated the search for a geometric description of backgrounds with fluxes, which was very much guided by the framework of generalized geometry developed by Hitchin \cite{Hitchin,Gualtieri}. In rough terms, generalized complex geometry is complex geometry applied to the generalized tangent bundle 
of the space, consisting of the sum of  tangent and cotangent bundles. The parameters encoding the symmetries of the metric plus the B-field, namely diffeomorphisms plus gauge transformations of B, live on this bundle. This formulation has therefore a natural action of T-duality, which exchanges these two. On the generalized tangent bundle one can define (generalized) almost complex structures, and study their integrability (integrable generalized complex structures allow to integrate the one-forms $dZ^i$ and find global complex coordinates). Generalized almost complex structures are in one-to-one correspondence with
pure spinors, which are built by tensoring the internal spinor with itself and with its charge conjugate. Spinors on the generalized tangent bundle are isomorphic to sums of forms on the cotangent bundle, and the integrability condition for the structure is nicely recast into closure of the pure spinor under the exterior derivative twisted by $H$\footnote{Integrability condition is actually weaker, it requires $(d -H \wedge) \Phi= X \Phi$ for some generalized tangent vector $X$, where $\Phi$ is the pure spinor corresponding to the generalized almost complex structure.}.    

Generalized complex geometry was used in  \cite{GMPT1,GMPT2} to  characterize $\mathcal{N}=1$ vacua. In analogy with the fluxless case, off-shell supersymmetry requires an algebraic condition, namely the existence of two pure spinors on the generalized tangent bundle. $\mathcal{N}=1$ vacua require one of the pure spinors to be closed (and therefore the generalized almost complex structure associated to be integrable), while RR fluxes act as a defect for integrability of the other structure. 
In order to geometrize the RR fields as well, and give a purely algebraic geometrical characterization of the vacua 
(which would allow, for example, to study their
deformations, i.e. their moduli spaces, in a model-independent manner), one needs to extend the generalized tangent bundle such that it includes the extra symmetries corresponding to gauge transformations of
the RR fields. Such extension has been worked out in \cite{Hull,PW}, and was termed exceptional generalized geometry, alluding to the exceptional groups arising in U-duality. In this paper we study compactifications of type II down to four-dimensions, where the relevant group is $\Es7$.

The algebraic conditions to have off-shell $\mathcal{N}=2$ supersymmetry in four-dimensions have been worked out in \cite{GLSW}. 
Very much in analogy to the generalized complex geometric case, they require the existence of two algebraic structures on the exceptional generalized tangent bundle (in fact one of them, rather than a single structure, is actually a triplet, satisfying an $SU(2)_R$ algebra), which are built by tensoring the internal spinors. The $SU(2)_R$-singlet structure, that we call $L$, describes the vector
multiplet moduli space, while the triplet of structures (named $K_a$) describes the hypermultiplets. The $\mathcal{N}=1$ 
preserved supersymmetry breaks the $SU(2)_R$ into $U(1)_R$, selecting a vector $r^a$ along this $U(1)$, and a complex orthogonal vector $z^a$. The complex combination $z^a K_a$ describes the $\mathcal{N}=1$ chiral multiplets contained
in the hypermultiplets\footnote{The vectors $r^a$ and $z^a$ are also used to identify respectively the $\mathcal{N}=1$ D term and superpotential out of the triplet of Killing prepotentials in $\mathcal{N}=2$ theories.}.

In this paper we obtain the differential conditions on the algebraic  structures $L, K_a$ required by  $\mathcal{N}=1$ on-shell supersymmetry\footnote{Steps in this direction were done in \cite{GLSW} (see also in \cite{PW} for the M-theory case), where a set of natural $\E7$-covariant equations was conjectured to describe $\mathcal{N}=1$ vacua. While the spinor components of such equations  reproduce those of \cite{GMPT2} and are therefore true conditions for susy vacua, other components failed to reproduce supersymmetry conditions.}. 
The first step is to identify the appropriate twisted derivative that generalizes $d-H \wedge$ to include the RR fluxes, or
in other words to identify the right generalized connection.  Such connection is obtained as in standard differential geometry by the operation $g^{-1} D g$, where $g$ are the $\E7$-adjoint elements
corresponding to the shift symmetries (the so-called "B- and C-transforms"), and the derivative operator $D$ is embedded in the fundamental
representation of $\E7$ \cite{GLSW}. The key point is that this connection, which a priori transforms as a generic tensor product of adjoint and fundamental representations, should only belong to a particular irreducible representation in this tensor product, which in the case at hand is the $\rep{912}$. 
Having identified the appropriate connection, we rewrite supersymmetry conditions in terms of closure of the structures.  
The equations we get are given in (\ref{N=1EGG1})-(\ref{vectorDK}). We find that  $\mathcal{N}=1$ supersymmetry 
requires on one hand closure of $L$, as conjectured in \cite{GLSW}. On the other hand, the components of the twisted derivative of $r^a K_a$ with an even number of internal indices  have to vanish, while those with an odd number are proportional to derivatives of the warp factor. A similar thing happens with $z^a K_a$, except that this time closure occurs upon projecting onto
the holomorphic sub-bundle defined by $L$.

The paper is organized as follows: in section \ref{geometry} we review the basic features of generalized geometry and its extension achieved by exceptional generalized geometry. In section \ref{lk} we present the relevant algebraic structures for compactifications with off-shell $\mathcal{N}=2$ supersymmetry. In section \ref{sintegrability} we review the constrains on the (traditional and generalized complex) structures imposed by on-shell supersymmetry. In section \ref{diffprev} we study supersymmetric vacua in the framework of exceptional generalized geometry. In particular, we introduce the twisted derivative operator in \ref{derivative},  we present the $\mathcal{N}=1$ equations in \ref{n1}, and finally in section \ref{n1proof} we outline the proof that supersymmetry requires those equations. Various technical details, as
well as the full derivation of the equations, are left to Appendices \ref{E7} to \ref{app:DLDKvssusy}.

\end{section}

\begin{section}{Generalized geometry}
\label{geometry}

\begin{subsection}{Generalized  Complex Geometry} \label{GCG}

In this section we present the basic concepts of Generalized Complex Geometry (GCG) in six-dimensions (we will restrict to the six-dimensional  case, though most of what follows can be generalized to any dimension), which will be used as a mathematical tool for describing flux vacua.   

In Generalized (Complex) Geometry, the algebraic structures are not defined on the usual tangent bundle $TM$ but on $TM\oplus T^*M$, on which there is a natural metric $\eta$ 
\begin{align}
\eta=\left(\begin{array}{cc}
0&1_6\\1_6&0\end{array}\right)\label{eta} \ .
\end{align}
Following the language of usual complex geometry, a generalized almost complex structure (GACS for short) $\cj$ is a map
from $TM\oplus T^*M$ to itself such that it satisfies the hermiticity condition ($\mathcal{J}^t\eta\mathcal{J}=\eta$) and $\mathcal{J}^2=-1_{12}$.
One can define projectors $\Pi_\pm$ for the complexified generalized tangent bundle as
\begin{equation}
\Pi_{\pm}=\frac{1}{2}(1_{12}  \mp i\mathcal{J})\label{proj}
\end{equation}
which can be used to define a maximally isotropic sub-bundle (six-dimensional) of $TM\oplus T^*M$ as the $i$-eigenbundle of $\cj$
\begin{equation}
L_{\mathcal{J}}=\{ x+\xi \in TM\oplus T^*M \big| \Pi_+ (x+\xi)=x+\xi \} \ .
\end{equation}
There is a one-to-one correspondence between a GACS and a ``pure spinor" $\Phi$ of $O(6,6)$. A spinor is said to be pure if its annihilator space
\begin{equation}
L_{\Phi}=\{x+\xi \in TM\oplus T^*M \big| (x+\xi)\cdot\Phi=0 \}
\end{equation}
is maximal (here $\cdot$ refers to the Clifford action $X \cdot \Phi=X_A \Gamma^A \Phi$). The one-to-one correspondence is then\footnote{The correspondence is actually one-to-many as the norm of the spinor is unfixed\label{foot:mto}.} 
\begin{equation}
\mathcal{J}\leftrightarrow \Phi, \mbox{ if } L_{\mathcal{J}}=L_{\Phi} \ .
\end{equation}
One can construct the GACS from the spinor by
\begin{equation}
\label{Jgen}
   \mathcal{J}^{\pm A}{}_B\ =\ i \, \frac{
       \mukai{\bar \Phi^\pm}{\Gamma^A{}_B {\Phi}^\pm}}
       {\mukai{\Phi^\pm}{\bar{\Phi}^\pm}}\ ,
\end{equation}
Weyl pure  spinors of $O(6,6)$ can be built by tensoring two $O(6)$ spinors $(\eta^1,\eta^2)$ as follows
\begin{equation}
\Phi^+=e^{-\phi} \eta^{1}_+ \eta^{2\dagger}_{+}, \qquad \Phi^-=e^{-\phi} \eta^{1}_+ \eta^{2\dagger}_{-}
\label{purespinors}
\end{equation}
 where the plus and minus refers to chirality, and $\phi$ is the dilaton, which defines
the isomorphism between the spinor bundle and the bundle of forms. Using 
Fierz identities, these can be expanded as
\begin{equation}
\eta^1_{\pm} \eta^{2\dagger}_{\pm}=\frac{1}{8}\overset{6}{\underset{k=0}\sum}\frac{1}{k!}(\eta^{2\dagger}_{\pm}\gamma_{m_k\dots i_1}\eta^1_{\pm})\gamma^{i_1\dots m_k} \ .
\label{bilinears}
\end{equation}
Using the isomorphism between the spinor bundle and the bundle of differential forms (often referred to as Clifford map):
\begin{equation}
 A_{m_1\dots m_k} \gamma^{m_1\dots m_k}  \longleftrightarrow A_{m_1 \dots m_k} dx^{m_1}\wedge\dots\wedge dx^{m_k}
\end{equation}
the spinor bilinears (\ref{bilinears}) can be mapped to a sum of forms. Under this isomorphism, the inner product of spinors  
$\Phi \chi$ is mapped to the following action on forms, called the Mukai pairing
\begin{equation} \label{Mukai}
\langle\Phi,\chi\rangle=(\Phi\wedge s(\chi))_6, \quad  \,\,\mbox{where     } s(\chi)=(-)^{\mbox{\begin{scriptsize}Int\end{scriptsize}}[n/2]}\chi
\end{equation}
and the subindex 6 means the six-form part of the wedge product.
 
For Weyl spinors, the corresponding forms are only even (odd) for
a positive (negative) chirality $O(6,6)$ spinor. 
In the special case where $\eta^1=\eta^2\equiv\eta$, familiar from Calabi-Yau compactifications, we get
\beq \label{Phipmsu3}
\Phi^+=e^{-\phi} e^{-iJ} \ , \qquad \Phi^-=-i e^{-\phi} \Omega
\eeq
where $J, \Omega$ are respectively the symplectic and complex structures of the manifold (more details are given in section 
\ref{susynofluxT}).  

Pure spinors can be ``rotated" by means of $O(6,6)$ transformations. Of particular interest is the nilpotent
subgroup of $O(6,6)$ defined by the generator
\beq \label{Btransform}
{\cal B}=  \begin{pmatrix}
            0 & 0 \\  B & 0        
            \end{pmatrix} \ , 
\eeq 
with $B$  an antisymmetric $6\times6$ matrix, or equivalently a two-form. On spinors, it amounts to the exponential action 
\beq \label{Btwistedspinors}
\Phi^\pm \to e^{-B} \Phi^\pm \equiv \Phid^\pm  
\eeq
where on the polyform associated to the spinor, the action is $e^{-B} \Phi=(1-B\wedge +\tfrac12 B \wedge B \wedge+...) \Phi$. We will refer to $\Phi$ as naked pure spinor, while $\Phid$ will be called dressed pure spinor.
The pair ($\Phid^+,\Phid^-$) defines a positive definite metric on the generalized tangent space,
which in turn defines a positive metric and a two-form (the $B$ field) on the six-dimensional manifold.

\end{subsection}

\begin{subsection}{Exceptional Generalized Geometry} \label{EGG}

Exceptional generalized geometry (EGG) \cite{Hull,PW} is an extension of the $O(6,6)$ (T-duality) covariant formalism of generalized geometry to an  $E_{7(7)}$ (U-duality) covariant one, such that the RR fields are incorporated into the geometry. 

We saw in the previous section that there is a particular $O(6,6)$ adjoint action (\ref{Btransform}) corresponding to shifts of the B-field. In EGG, shifts of the B-field as well as shifts of the sum of internal RR fields $C^{-}=C_{1}+C_{3}+C_{5}$ \footnote{In this paper we will concentrate on type IIA, but most of the statements can be easily changed to type IIB by switching chiralities.}, which transforms as a chiral $O(6,6)$ spinor, correspond to particular $\E7$ adjoint actions. To form a set of gauge fields that is closed under U-duality, we also have to consider the shift of the six-form dual to $B_2$, which we will call $\tilde B$.\footnote{Equivalently these are shifts of the dual axion $B_{\mu\nu}$.}  

Decomposing the adjoint  \textbf{133} representation of $E_{7(7)}$ under $\Tsub$, we have 
\begin{align}
 \label{adjoint}
 \bf{133}&=\bf{(3,1)}+\bf{(1,66)}+\bf{(2,32')} \\
            \mu&=(\mu^{i}_{\,\,\,j} \ , \ \mu^{A}_{\,\,\,B}\ ,\ \mu^{i-})\nonumber
\end{align}
where $i=1, 2$ is a doublet index of $\SLR$, raised and lowered with $\epsilon_{ij}$, and the $O(6,6)$ fundamental indices $A, B=1,...,12$ are raised and lowered with the 
metric ${\eta}$ in (\ref{eta}). The B-transform action (\ref{Btransform}) is part of $\mu^A{}_B$, while the C-transformations are    
naturally embedded in one of the two $\bf{32'}$ representations. Let us call $v^i$ the $\SLR$ vector pointing in the direction 
of the C-field, which we can take without loss of generality to be 
\beq
v^i=(1,0) \ .
\eeq
The $GL(6)$ assignments of the different components shown in Appendix \ref{switch}, indicate that the shift symmetries are given by the following sum of generators
\beq \label{calA}
\left( \tilde B v^{i}v_{j}, \left(\begin{array}{cc} 0&0\\B&0\end{array}\right), v^{i}C^{-} \right) \equiv {\cA} \ 
\eeq
where $v_i=\epsilon_{ij} v^j$. Using (\ref{adjac}) it is not hard to show that given this embedding we recover the right commutation relations
\begin{equation}
\label{form-alg}
   \big[B + \tilde B + C^- , B' +\tilde B' +  C^{-\prime} \big]
      = 2\langle C^- , C^{-\prime}\rangle 
          + B \wedge C^{-\prime} - B' \wedge C^- \ , 
\end{equation}
where the first term on the rhs is a six-form and therefore corresponds to a $\tilde B$ transformation and the other two, to an RR shift.  

The fundamental \rep{56} representation of $\E7$ decomposes under $\Tsub$ as
\begin{align}
\bf{56}&=\bf{(2,12)}+\bf{(1,32)}\label{fundgeom} \\
\nu&=(\nu^{iA},\nu^{+})\nonumber \ .
\end{align}
It combines all the gauge transformations: vectors plus one-forms correspond to diffeomorphisms and gauge transformations of the B-field. Their $\SLR$ duals\footnote{The $\SLR$ here is the ``heterotic S-duality", where the complex field that transforms
by fractional linear transformations is $S=\tilde{B}+i.e.^{-2\phi}$. For the connection between this and type IIB S-duality, see \cite{AAGP}.} are gauge transformations of $B_6$ (given by a five-form, or analogously a vector) and diffeomorphisms for the dual vielbein (sourced by KK monopoles), given by a one-form tensored a six-form. Gauge transformations of the RR fields combine forming again a spinor representation, this time with positive chirality. The generalized tangent bundle $T \oplus T^*$ 
is therefore extended to the exceptional tangent bundle (EGT) $E$
 \beq \label{fund}
 E=TM\oplus T^*M \oplus   \Lambda^5T^*M\oplus (T^*M\otimes \Lambda^6T^*M)\oplus\Lambda^{\mbox{\begin{scriptsize}even\end{scriptsize}}}T^*M \ .
\eeq 

In what follows, we will mostly use the decomposition of $\E7$ under $\SLE$. The fundamental representation decomposes as
\begin{align}
\bf{56}&=\bf{28}+{\bf{28'}} \label{slf} \quad \\
\nu&=(\nu^{ab}, \tilde{\nu}_{ab})\nonumber \ 
\end{align}
where $a,b=1,...,8$ and $\nu_{ab}=-\nu_{ba}$. The adjoint decomposes as 
\begin{align}
\bf{133}&=\bf{63}+\bf{70}\label{sla} \\
\mu&=(\mu^{a}_{\,\,\,b},\mu_{abcd}) \nonumber
\end{align}
where $\mu^a{}_a=0$ and $\mu_{abcd}$ is fully antisymmetric.

In order to identify the embedding of the gauge fields (\ref{calA}) in $\SLE$, we use the $GL(6,\mathbb{R})$ properties of the 
different components of the adjoint representation given in (\ref{musl8prop}).  We get \footnote{To avoid introducing new notation, we are using the same as in (\ref{calA}), in particular $v_i\equiv \epsilon_{ij} v^j$, although indices in $\SLE$ are raised and lowered with the metric $\hat g$ given in (\ref{SLEmetric}).} 
\beq \label{calAsl81}
{\cA}=\left(e^{2\phi} \tilde B v^i v_j -v^i e^{\phi} C_m  + e^{\phi} (*C_{5})^m  v_i \,,    -\tfrac12 e^{\phi} C_{mnp} v_i -\tfrac12 B_{mn} \epsilon_{ij} \right) \ ,
\eeq
or in other words
\begin{align} \label{calAsl8}
\cA^1{}_2&=-e^{2\phi} \tilde B \ , \quad \cA^1{}_m=-e^{\phi} C_m \ , \quad \cA^m{}_2=-e^{\phi} (*C_{5})^m \nn \\
\cA_{mnp2}&=\tfrac12 e^{\phi} C_{mnp} \ , \quad  \cA_{mn12}=-\tfrac12 B_{mn}
\end{align}
where the factors and signs are chosen in order to match the supergravity conventions. Here and in the following, $*$ refers to a six-dimensional Hodge dual, while we use $\star$ for the eight-dimensional one.

\end{subsection}
\end{section}

\begin{section}{$E_{7(7)}$ algebraic structures}\label{lk}

In this section we present the algebraic structures in $\E7$ constructed in \cite{GLSW} that play the role of the $O(6,6)$ pure spinors $\Phi^\pm$. We start by building the analogous of the naked pure spinors, and then discuss their orbits under the action of the gauge fields ${\cA}$ in (\ref{calA}), (\ref{calAsl81}).

Spinors transform under the maximal compact subgroup of the duality group. In the GCG case, this subgroup is $O(6) \times O(6)$, which acts on the pair $(\eta^1,\eta^2)$. In EGG, the relevant group is $SU(8)$.
We can combine the two ten-dimensional supersymmetry parameters such that the $SU(8)$ transformation
of their internal piece is manifest. The most general ten-dimensional spinor ansatz relevant to four-dimensional $\mathcal{N}=2$    
theories is
\begin{align}
\left(\begin{array}{c}\epsilon^1\\ \epsilon^2\end{array}\right)=\zeta^1_-\otimes \theta^1 +\zeta_-^2\otimes \theta^2 +\mbox{c.c.}\label{suvar}
\end{align}
where $\zeta^{1,2}_-$ are four-dimensional spinors of negative chirality, and $\theta^{1,2}$ are never parallel.
In this paper we will be dealing with equations for ${\mathcal N}=1$ vacua, where there is a relation between $\zeta^1$ and $\zeta^2$. In that case, we can use the special parameterization
\begin{equation}
\theta^1=\left(\begin{array}{c} 
\eta^{1}_{+}\\
0
\end{array}\right), \qquad
\theta^2=\left(\begin{array}{c} 0 \\
\eta^2_{-} 
\end{array}\right)\label{fulla} \ .
\end{equation}
A nowhere vanishing spinor $\theta$ defines an $SU(7)\subset SU(8)$ structure. The pair $(\theta^1,\theta^2)$ defines an $SU(6)$ structure\footnote{Note that an $SU(6)$ structure can be built out of  a single globally defined internal spinor $\eta$, taking  $\eta^1=\eta^2=\eta$.}. 
We can take the $SU(4)$ spinors to be normalized to $1$. In that case the $SU(8)$ spinors are orthonormal, namely 
\beq \label{normtheta}
\bar \theta_I \, \theta^J= \delta_I{}^J \ .
\eeq 
where $I=1,2$ is a fundamental $SU(2)_R$ index (for conventions on the conjugate spinors, see Appendix \ref{gamma}). 
 The two spinors can be combined into the following
$SU(2)_R$ singlet and triplet combinations
\beq \label{lambdaK}
L=e^{-\phi} \epsilon_{IJ}\theta^{I}\theta^{J} \ , \qquad K_{a} =\frac{1}{2} e^{-\phi} \sigma_{aI}{}^J\theta^{I}\bar{\theta}_{J} \ ,  \qquad K_{0} =\frac{1}{2} e^{-\phi} \delta_{I}{}^J\theta^{I}\bar{\theta}_{J} \ ,
\eeq
where  we have introduced $K_0$ for future convenience. The triplet
$K_a$ satisfies the $su(2)$ algebra with a scaling given by the dilaton, i.e.
\beq
[K_a , K_b]= 2i  e^{-\phi} \epsilon_{abc} K_c
\eeq
$L$ and $K_a$ are the $\E7$ structures that play the role of the generalized almost complex structures $\Phi^+$ and $\Phi^-$.
They belong respectively to the $\rep{28}$ and $\rep{63}$ representations of $SU(8)$, which are in turn part of the $\rep{56}$ and $\rep{133}$ representations of $\E7$. Using the decompositions $\rep{56}=\rep{28} + \overline{\rep{28}}$ and $\rep{133}=\rep{63}+\rep{35}+\overline{\rep{35}}$ shown in (\ref{suf}) and (\ref{sua}), they read
\beq \label{lambdaKsu8}
L=\left(e^{-\phi} \epsilon_{IJ}\theta^{I \alpha}\theta^{J \beta}, e^{-\phi} \epsilon_{IJ}\theta^{I*}_ {\alpha}\theta^{J*}_ {\beta} \right) \, \qquad K_{a}=\left(e^{-\phi} \tfrac{1}{2}\sigma_{aI}{}^J\theta^{I \alpha}\bar{\theta}_{J \beta},0,0 \right) \ .
\eeq

To make contact with the pure spinors of GCG, we note that using the parameterization (\ref{fulla}), we get
\begin{align} \label{Lstt}
L=\left(\begin{array}{cc} 0 & \Phi^{+} \\
-s(\bar{\Phi}^{+}) & 0
\end{array} \right)
\end{align}
where the operation $s$ is introduced in (\ref{Mukai}).

Using (\ref{fulla}), we get for $K_\pm=K_1 \pm i K_2$
\begin{equation} \label{Kstt}
K_{+}=\left(\begin{array}{cc}
0 & \Phi^{-}\\
0 & 0
\end{array}\right) \ , \qquad
K_{-}=\left(\begin{array}{cc}
0 & 0 \\
-s(\bar{\Phi}^{-}) & 0
\end{array}\right) \ , 
\end{equation}
while for $K_3$ we get
\beq
K_{3}=\left(\begin{array}{cc}
\Phi_1^+ & 0 \\
0 &  - \bar{\Phi}_2^+ 
\end{array}\right) \nn
\eeq
where we have defined 
\beq
\Phi^+_1  = e^{-\phi} \eta^1_+ \eta^{1 \dagger}_+ \ , \qquad \Phi^+_2  =e^{-\phi}  \eta^2_+ \eta^{2 \dagger}_+ \ , 
\eeq  
We see that $L$ contains the pure spinor $\Phi^+$, which spans the vector multiplets in type IIA (see (\ref{Phipmsu3})),
while $K_+$ is built from the pure spinor $\Phi^-$, which is part of the hypermultiplets. 
$K_3$ contains on the contrary the even-form bilinears of the same $SU(4)$ spinor, or in other terms the symplectic structures defined by each spinor (see (\ref{Phipmsu3})). 

To get the $SL(8)$ components of $L$ and $K_a$, we use (\ref{pw2}).  Using the decomposition of the gamma matrices
given in (\ref{gammabasis}), we get that the only non-zero components of $L$ and $K_a$ are
\begin{align} \label{nonzerocomp}
L&: \qquad L^{12}, L^{mn} \nn\\
K_1, K_2&: \qquad K_{2}{}^{m1},  K_2{}^{m2},  K_2{}^{mnp1},  K_2{}^{mnp2}  \\
K_0, K_3&: \qquad K_3{}^{mn},  K_3{}^{12},  K_3{}^{mnpq},  K_3{}^{mn12} \nn
\end{align}
where $L^{12}$ and $L^{mn}$ involve the zero and two-form pieces of $\Phi^+$, $K_{+}^{mi}, K_{+}^{mnpi}$ contain the one and three-form pieces of $\Phi^+$
(where the difference between the two $SL(2)$ components is a different $GL(6)$ weight), while $K_3$ contains the different components of $\Phi^+_1$ and $\Phi^+_2$.

In an analogous way as for the pure spinors, the structures $L$ and $K_{a}$ 
can be dressed by the action of the gauge fields $B$, $\tilde B$ and $C^-$ in (\ref{calA}), (\ref{calAsl8}), i.e. we define
\beq
\label{fulllambdaK}
L_D=e^{C} e^{\tilde B} e^{-B}  L \ , \qquad K_{aD}=e^{C} e^{\tilde B} e^{-B}  K_{a} \ .
\eeq

In the GCG case, the B-field twisted pure spinors span the orbit $\frac{O(6,6)}{SU(3,3)} \times {\mathbb R}^+$, where $SU(3,3)$ is the stabilizer of the pure spinor and the ${\mathbb R}^+$ factor corresponds to the norm. Quotenting by the ${\mathbb C}^*$ action $\Phid \to c \Phid$, we get the space $\frac{O(6,6)}{U(3,3)}$ which is local Special K\"ahler. 
Similarly, our EGG structures $L_D$ and $K_{aD}$ span orbits in $\E7$ which are respectively Special K\"ahler and
Quaternionic-K\"ahler. As shown in \cite{GLSW}, the structure $L_D$ is stabilized by $\Ex6$, and the corresponding local 
Special K\"ahler space is $\frac{\E7}{\Ex6} \times U(1)$. 
%
The triplet $K_{aD}$ is stabilized by an $SO^*(12)$
subgroup of $\E7$, and the corresponding orbit is the quaternionic space $\frac{\E7}{SO^*(12) \times SU(2)}$, where
the $SU(2)$ factor corresponds to rotations of the triplet. The $SO^*(12)$ and $\Ex6$ structures intersect on an
$SU(6)$ structure if $L$ and $K_a$ satisfy the compatibility condition
\beq
L \, K_a|_{\bf 56}=0 \ ,
\eeq
where we have to apply the projection on the $\rep{56}$ on the product $\rep{56} \times \rep{133}$. This condition is automatically satisfied for the structures (\ref{lambdaK}) built  as spinor
bilinears.

\end{section}

\begin{section}{String vacua and integrability conditions}\label{sintegrability}

In the previous sections we have presented the relevant algebraic structures that are used to describe
an off-shell $\mathcal{N}=2$ four-dimensional effective action. We now turn to the differential conditions imposed by
requiring on-shell supersymmetry, or in other words, by demanding that the vacua are supersymmetric. 
As we will show, these translate into integrability of some of the algebraic structures.

\begin{subsection}{Warm up: fluxless case}

It will be useful for the following to recall the conditions for supersymmetric vacua in the absence of fluxes.
We start by reviewing the integrability conditions in ordinary complex geometry, and then re-express them
in the language of GCG.

\begin{subsubsection}{Conditions for the structures on $TM$}
\label{susynofluxT}

In the absence of fluxes, inserting the $\mathcal{N}=2$ spinor ansatz (\ref{fulla}) in the supersymmetry condition $\delta \psi_m=0$ (see (\ref{svariations2})), we get 
\beq \label{nablatheta0}
\nabla_m \theta^I=0 \ .
\eeq
When there is only one globally defined spinor $\eta$, we take  $\eta^1=\eta^2\equiv \eta$, and 
(\ref{nablatheta0}) reduces to the familiar Calabi-Yau condition
\begin{equation} \label{nablaeta0}
\nabla_m \eta=0 \ ,
\end{equation}
which implies that the $SU(3)$ structure defined by $\eta$ is integrable, or in other words that the manifold has $SU(3)$ holonomy \cite{joyce}. The holonomy is defined as the group generated by parallel transporting an arbitrary spinor around a closed loop.  Riemaniann geometries can be classified by specifying the holonomy of the Levi-Civita connection. A general Riemaniann six-dimensional space has holonomy $SO(6)\simeq SU(4)$. However if the manifold admits one (or more) Killing spinors, the group is reduced:  it lies within the stabilizer group. In six dimensions, the existence of a globally defined, nowhere vanishing, covariantly constant spinor implies that the holonomy is reduced to $SU(3) \subset SU(4)$. 

Integrability of an $SU(3)$ structure can also be recast in terms of integrability of two seemingly very different algebraic structures that intersect on an $SU(3)$, namely a complex and a symplectic one. The existence of a globally defined nowhere vanishing spinor is equivalent to the existence of an almost symplectic 2-form $J$ (which defines an almost symplectic $Sp(6,{\mathbb R})$ structure) and a 3-form $\Omega$ (which defines an almost complex $GL(3,\mathbb{C})$ structure).
These two structures intersect on an $SU(3)$. If the structures are integrable, i.e. if they satisfy 
\begin{equation}
 dJ=0 \ , \qquad\,\,\, d\Omega =\xi \wedge \Omega \label{integrability} \ ,
\end{equation}
for any one-form $\xi$, one can define local complex and local symplectic coordinates which can be ``integrated" (i.e.  there exist local complex coordinates $z^i$ and symplectic ones $(x^i, y^{\hat \imath})$ ($i,\hat \imath=1,2,3$) such that the local complex and symplectic one forms $dz^i$, $(dx^i,dy^{\hat \imath})$ are indeed their differentials). If additionally $\xi=0$, then the canonical bundle is holomorphically trivial and the manifold is Calabi-Yau. Since $J$ and $\Omega$ can be written as bilinears of the spinor $\eta$, the supersymmetry requirement (\ref{nablaeta0}) is equivalent to the conditions (\ref{integrability}) and the additional requirement $\xi=0$. 

Note that for an almost complex structure, there are many equivalent ways to check its integrability. Instead of the second requirement in (\ref{integrability}), one can find conditions on the corresponding map $I: TM \to TM$\footnote{Similarly to the case of GACS, there is a one-to-one (or rather many-to-one (see footnote \ref{foot:mto})) correspondence between a 3-form $\Omega=dz^1\wedge dz^2\wedge dz^3$ and a map $I$ satisfying $I^2=-1$ such that the $i$-eigenbundle of $I$ is generated by the dual vectors $\partial_{z^i}$.}. The almost complex structure $I$ is integrable if the $i$-eigenbundle is closed under the Lie bracket, i.e. iff
\begin{equation} \label{intbracket}
\pi_{\mp}[\pi_{\pm} \, x ,\pi_{\pm} \, y ]=0, \quad \forall \, x, y \in TM \ \qquad {\rm where} \ \pi_{\pm}=\frac{1}{2}(1\mp iI)
\end{equation}
and $[ \ , \ ]$ denotes the Lie bracket. As we will see, either requirement (\ref{integrability}) and (\ref{intbracket}) will have its analogue in generalized complex geometry. In exceptional generalized geometry, we will only deal with conditions of the form (\ref{integrability}).

\end{subsubsection}

\begin{subsubsection}{Conditions for the structures on $TM\oplus T^*M$}

As shown in section \ref{GCG}, almost complex and symplectic structures on the tangent bundle are expressed on the same footing in terms of generalized almost complex structures on $TM\oplus T^*M$. Furthermore, a generic GACS reduces on the tangent bundle to a structure that is locally a product of lower dimensional complex and symplectic structures. 

As in the case of ordinary complex structures, Eq.(\ref{intbracket}), a GACS is integrable if its $i$-eigenbundle is closed under 
an extension of the Lie bracket to $T\oplus T^*$, i.e. ${\cal J}$ is integrable iff
\begin{equation} \label{intbracketGACS}
\Pi_{\mp}[\Pi_{\pm}(X),\Pi_{\pm}(Y)]_C=0,\quad \forall X, Y \in TM\oplus T^*M 
\end{equation}
where the projectors $\Pi_\pm$ are defined in (\ref{proj}) and the bracket is the Courant bracket
\begin{equation} \label{Courant}
[x+\xi,y+\eta]_C=[x,y]+\mathcal{L}_x\eta-\mathcal{L}_y\xi-\frac{1}{2}d(i_x\eta-i_y\xi)
\end{equation}
with ${\cal L}$ the Lie derivative. Again, in a similar fashion to ordinary complex structures, the integrability condition 
(\ref{intbracketGACS}) is equivalent to requiring that the pure spinor $\Phi$ associated to ${\cal J}$ satisfies
\beq
d\Phi=X \cdot \Phi
\eeq
for some generalized vector  $X=x+\xi$, and where $\cdot$ is the Clifford product, whose action on forms is
\begin{equation}
X \cdot\Phi=\iota_{x}\Phi+\xi\wedge\Phi \label{caction} \ .
\end{equation}

The $\mathcal{N}=2$ supersymmetry requirement (\ref{nablatheta0}) that arises in the absence of fluxes, translates into 
\beq \label{GCYM}
d\Phi^+=0 \ , \qquad d\Phi^-=0 \ ,
\eeq 
which means that both GACS are integrable (and both canonical bundles are trivial), or in other words that the $SU(3) \times SU(3)$ structure is integrable. In the case $\eta^1=\eta^2=\eta$, this reduces to the Calabi-Yau conditions
(\ref{integrability}) with $\xi=0$. Manifolds satisfying (\ref{GCYM}) have been termed  ``generalized Calabi-Yau {\it metric} geometries" in \cite{Gualtieri}\footnote{\label{foot:GCY}Note the addition of the word ``metric", to distinguish them from the generalized Calabi-Yau manifolds defined in \cite{Hitchin} that require closure of only one pure spinor, and will play a main role in the next sections.}. They are more general than Calabi-Yau's in the sense that the pure spinors need not be purely complex or pure symplectic, as happens when $\eta^1=\eta^2$, but can correspond to (integrable) hybrid complex-symplectic structures.

\end{subsubsection}
\end{subsection}

\begin{subsection}{Flux case in CGC}

In this section we review the results of \cite{GHR} (in the language of GCG, as in \cite{JW}) and \cite{GMPT2} where the conditions for respectively $\mathcal{N}=2$ supersymmetry with NS flux only, and $\mathcal{N}=1$ with NS and RR fluxes were found.

\begin{subsubsection}{Vacua with NS fluxes}

In section \ref{GCG} we saw how GCG incorporates the B-field, in particular by means of the B-twisted pure spinors
(\ref{Btwistedspinors}). When $B$ is not globally well-defined, i.e. when NS fluxes are switched on, the B-twisted pure spinors 
are not global sections of $TM\oplus T^*M$, but they are rather sections of a particular fibration of $T^*M$ over $TM$
involving the $B$-field. For reasons that will become clear later, in this paper we choose the alternative ``untwisted picture"   
as in \cite{Gualtieri}, where pure spinors are naked (or dressed by just a closed $B$ field), and the $H$-flux is introduced explicitly in, e.g. the integrability conditions\footnote{\label{foot:twisted}We use the terming ``twisted picture" to refer to the scenario where pure spinors are dressed by the (non-closed) $B$-field, and the integrability conditions are given in terms of the ordinary exterior derivative (or equivalently the ordinary Courant bracket (\ref{Courant})), as in \cite{Hitchin},  while in the ``untwisted picture" of \cite{Gualtieri}, the spinors are untwisted (or just twisted by a closed $B$), while the $H$-flux
appears explicitly in the differential or in the bracket. The two pictures are equivalent, and depending on the situation one can be more convenient than the other.}.  

A closed $B$ field is an automorphism of the Courant bracket, while in the presence of $H=dB$ flux, there is an extra term
\begin{equation}
[e^{-B}(x+\xi),e^{-B}(y+\eta)]_C=e^{-B}[x+\xi,y+\eta]_C + e^{-B}\iota_x\iota_y H \label{twist} 
\end{equation}
where the action of $B$ is
$
e^{-B}(x+\xi)=x+\xi-\iota_xB \ .
$
The $H$-twisted Courant bracket is defined by adding this last term to (\ref{Courant}). 

If a GACS is ``twisted integrable", then the corresponding pure spinor satisfies
\beq
d_H \Phi=X \cdot \Phi
\eeq
where the $H$-twisted differential is 
\begin{equation}
d_H\equiv d-H\wedge \ .
\end{equation}
Note the equivalence between the twisted and untwisted picture. If a naked pure spinor is twisted closed, then
the dressed pure spinor is closed under the ordinary exterior derivative, i.e.
\beq
0=d_H \Phi =(d-dB \wedge) \Phi = e^B d(e^{-B} \Phi) =e^B d \Phid \ .
\eeq
This shows how to construct the twisted exterior derivative from the ordinary one, and the action of the $B$-field
\beq \label{dtwistB}
d_H=e^B d e^{-B} 
\eeq
which will be extended in section (\ref{derivative}) to include the RR fluxes.

Supersymmetry conditions in the presence of $H$-flux amount precisely to $H$-twisting the generalized Calabi-Yau metric condition (\ref{GCYM}). More precisely,
vacua preserving four-dimensional ${\mathcal N}=2$ supersymmetry in the presence of NS fluxes should satisfy \cite{JW}
\beq \label{GCYMphi}
d_H \Phi^+=0 \ , \qquad d_H \Phi^-=0 \ ,
\eeq 
i.e. they require $H$-twisted generalized Calabi-Yau metric structures. 

Vacua with $\mathcal{N}=1$ supersymmetry in the presence of NS fluxes were obtained in \cite{torsion}, and reinterpreted in the language of G-structures in \cite{Gstructures}. They read
\begin{align}
d_H(e^{-\phi} \Phi^-)&=0 \ , \nn \\
 d(e^{-\phi} \Phi^+) &=i e^{-2\phi} * H
\end{align}
where $\Phi^\pm$ are those for an $SU(3)$ structure, (\ref{Phipmsu3}).  Note that
in the second equation $H$ does not enter as a twisting in the standard way, and therefore the even pure spinor is not twisted integrable.
It would be interesting to get the right GCG description of $\mathcal{N}=1$ vacua with NS fluxes.

\end{subsubsection}

\begin{subsubsection}{Vacua with NS and RR fluxes} \label{sec:N12GCG} 

Compactifications on Minkowski space preserving $\mathcal{N}=1$ supersymmetry in the presence of NS and RR fluxes require 
the spacetime to be a warped product, i.e.
\beq \label{warped}
ds^2=e^{2A} \eta_{\mu\nu} dx^\mu dx^\nu + ds_6^2 \ .
\eeq
The preserved spinor can be parameterized
within the $\mathcal{N}=2$ spinor ansatz (\ref{fulla}) by a doublet $n_I=( a, \bar b)$
such that the supersymmetry preserved is given by $\epsilon=n_I \epsilon^I$, i.e. 
\beq \label{thetaN=1}
\epsilon=\xi_- \otimes \theta + c.c. \ , \qquad {\rm with} \ 
\theta= \left(\begin{array}{c} a \eta^1_+ \\ \bar b \eta^2_{-} \end{array}\right)       \ ,
\eeq
and we take $|\eta^1|^2=|\eta^2|^2=1$ (while $|a|$ and $|b|$ are related to the warp factor, as we will see).
The vector $n_I$ distinguishes a $U(1)_R \subset SU(2)_R$ such that any triplet can be written in terms of a 
$U(1)$ complex doublet and a $U(1)$ singlet by means of the vectors 
\begin{align} \label{zr}
(z^+,z^-,z^3)&=n_I (\sigma^a)^{IJ} n_J=(a^2,-\bar b^2, -2a\bar b) \ , \\
(r^+,r^-,r^3)&=n_I  (\sigma^a)^I{}_J \bar n^J=(ab,\bar a \bar b, |a|^2-|b|^2) \nn \ .
\end{align}
Using these vectors, one can extract respectively an $\mathcal{N}=1$ superpotential and D-term from the triplet
of Killing prepotentials ${\cal P}_a$ that give the potential in the $\mathcal{N}=2$ theory, by
\beq
 {\cal W}=z^a {\cal P}_a \ , \qquad
{\cal D}=r^a  {\cal P}_a \ .
\eeq
 For type IIA compactifications, the triplet ${\cal P}_a$ reads \cite{GLW}
\beq \label{Pa}
{\cal P}_+=\langle \Phi^+, d_H \Phi^- \rangle \ , \quad {\cal P}_-=\langle \Phi^+, d_H \bar \Phi^- \rangle \ , \quad
{\cal P}_3=-\langle \Phi^+, F^+ \rangle \ .
 \eeq

The conditions for flux vacua have been
obtained in the language of GCG either using the ten-dimensional gravitino and dilatino variations \cite{GMPT2}, or by extremizing the superpotential of the four-dimensional $\mathcal{N}=1$ theory and setting the D-term  to zero  \cite{KM,BC}. For the case $|a|=|b|$, which arises when sources are present, they read
\begin{align} 
d_H(e^{2A}\Phi'^{+})&=0\label{koerber1}\\
d_H(e^{A}\mbox{Re}\Phi'^{-})&=0 \label{koerber2}\\
d_H( e^{3A} \mbox{Im}\Phi'^-)&=e^{4A} *s(F^+) \label{koerber3}
\end{align}
where 
\beq \label{phiprime}
\Phi'^+=2 a\bar b \, \Phi^+ \ , \quad \Phi'^-=2 a b \, \Phi^- \ .
\eeq
Finally, $N=1$ supersymmetry requires
\beq \label{norma}
|a|^2+|b|^2=e^{A} \ .
\eeq 
Conditions (\ref{koerber1})-(\ref{koerber3}) can be understood as coming from F and D-term equations. Equation (\ref{koerber2}) corresponds to imposing ${\cal D}=0$, while (\ref{koerber1}) and (\ref{koerber3})  come respectively from variations of the superpotential with respect to $\Phi^-$ and $\Phi^+$. 

 The susy condition (\ref{koerber1}) says that the GACS corresponding ${\cal J}^+$ is twisted integrable, and furthermore that the canonical bundle is trivial, and therefore the required manifold is a twisted Generalized Calabi-Yau (see footnote \ref{foot:GCY}). The other GACS appearing in (\ref{koerber2})-(\ref{koerber3})  is ``half integrable", i.e. its real part is closed, while the non-integrability of the imaginary part 
is due to the RR fluxes. In the EGG formulation, RR fluxes are also encoded in the twisting of the differential operator, and therefore we expect to rephrase these equations purely in terms of integrability of the structures defined on the EGT space. 
Note that in the limit of RR fluxes going to zero, Eqs. (\ref{koerber1})-(\ref{koerber3}) for $\mathcal{N}=1$ vacua reduce to (\ref{GCYMphi}) (for $F=0$, (\ref{koerber1})-(\ref{koerber3}) imply $A=0$), i.e. $F\to 0$ is a singular limit of (\ref{koerber1}) where supersymmetry is enhanced to $\mathcal{N}=2$.

On top of supersymmetry conditions (\ref{koerber1})-(\ref{koerber3}), the fluxes must satisfy the Bianchi identities
\beq
dH=0 \ , \quad d_H F =0
\eeq
in the absence of sources, while in the presence of D-branes or orientifold planes, the right hand sides get modified by the appropriate charge densities.

\end{subsubsection}

\end{subsection}
\end{section}

\begin{section}{Flux vacua in Exceptional Generalized Geometry}\label{diffprev}

In this section we discuss the conditions for $\mathcal{N}=1$ vacua in the language of EGG. The putative conditions
for supersymmetric vacua come from variations of the $\E7$-covariant expression for the triplet of Killing prepotentials \cite{GLSW}
\beq \label{Killprep}
{\cal P}_a={\cal S}(L_D, D K_{aD})={\cal S}(L, e^B e^{-\tilde B} e^{-C} D e^{C} e^{\tilde B} e^{-B} K_{a}) \ .
\eeq
Here ${\cal S}$ is the symplectic invariant on the $\rep{56}$ whose decomposition in terms of $\Tsub$ and $\SLE$ are given
respectively in (\ref{symplTsub}) and (\ref{symplSLE}). The derivative $D$ is an element in the $\rep{56}$, whose $\Tsub$ decomposition is
\beq \label{LC}
D=(D^{iA},D^+)=(v^i \nabla^A,0) \ , \qquad {\rm where} \ \nabla^A=(0,\nabla_m) \ ,
\eeq
while in $\SLE$ we have
\beq \label{LCsl8}
D=(D^{ab},\tilde D_{ab})=(0, v_i \nabla_m )  \ .
\eeq
(where we are using again $v_i=\epsilon_{ij} v^j=(0,-1)$), $D K_a$ in (\ref{Killprep}) is an element in the $\rep{56} \times \rep{133}$, which
is projected to the $\rep{56}$ by the symplectic product. In the second equality in (\ref{Killprep}) we have used the $\E7$ invariance
of the symplectic product to untwist the structures $L_D$ and $K_{aD}$ and express the Killing prepotentials
in terms of naked structures, and a twisted derivative. We will now see how to properly define this twisted derivative,
needed to get the equations for vacua.

\begin{subsection}{Twisted derivative and generalized connection}\label{derivative}

For the gauge fields ${\cA}$ and the derivative operator $D^{\cal A}$, ${\cal A}=1,...,56$, one can define a connection $\phi^{\cal AB}{}_{\cal C} \in \bf{56}\times\bf{133}$ by the following twisting of the Levi-Civita one
\beq
(e^B e^{-\tilde B} e^{-C})^{{\cal B}}{}_{\cal D} D^{\cal A} (e^{C} e^{\tilde B} e^{-B})^{\cal D}{}_{\cal C} \equiv D^{\cal A} \delta^{\cal B}{}_{\cal C} + \phi^{\cal AB}{}_{\cal C}\ .
\eeq 
The connection $\phi$ contains derivatives of the gauge fields. The key point is that in the tensor product
\begin{equation}
\bf{56}\times\bf{133}=\bf{56}+\bf{912}+\bf{6480}\label{fsot}
\end{equation}
only the terms in the ${\bf 912}$ representation involve exterior derivatives of the gauge potentials \cite{Udual}, while the other representations contain non-gauge invariant terms (like divergences of potentials).  
We therefore define the twisted derivative as
\beq\label{calD}
{\cal D}=D+ {\cal F} \ , \quad {\rm where} \ {\cal F}=e^B e^{-\tilde B} e^{-C} D \, e^{C} e^{\tilde B} e^{-B}\big|_{\bf{912}} \ .
\eeq
The fact that the fluxes lie purely in the $\rep{912}$ is consistent with the supersymmetry requirement that the embedding tensor of
the resulting four-dimensional gauge supergravity be in the $\rep{912}$ \cite{STW}. 

The $\rep{912}$ decomposes in the following $\Tsub$ representations
\begin{align}
\cf&=(\cf^{iA},\cf^{i}{}_{j}{}^+,\cf^{A-},\cf^{iABC}) \nn \\
\bf{912}&=\bf{(2,12)}+\bf{(3,32)}+\bf{(1,352)}+\bf{(2,220)} \nn
\end{align}
where $\Gamma_A \cf^{A-}=0$ and $\cf^{iABC}$ is fully antisymmetric in $ABC$. The only nonzero components
of the connection (\ref{calD}) are (see Appendix \ref{app:derivative} for details)
\beq \label{TsubF}
{\cal F}^{1}{}_{2}{}^+=-F^+ \ , \qquad {\cal F}^1{}_{mnp}=-H_{mnp} \ ,
\eeq
where $F^+=e^B dC^-$. 

In the $\SLE$ decomposition, the generalized connection decomposes in the following representations
\begin{align}
\bf{912}&=\bf{36}+\bf{420}+\bf{36'}+\bf{420'}\label{sl912} \\
\cf &= (\cf^{ab},\cf^{abc}{}_{d},\tilde \cf_{ab},\tilde \cf_{abc}{}^{d}) \nonumber
\end{align}
where $\cf^{ba}=\cf^{ab}$ and $\cf^{abc}{}_c=0$ and similarly for the objects with a tilde. The NS and RR fluxes give the following non-zero components
\begin{align} \label{sl8F}
\cf^{11}&= e^{\phi} \, F_0 \ , \qquad \cf^{mnp}{}_{2}=-\frac{1}{2}(*H)^{mnp} \ , \qquad
\cf^{mn1}{}_{2}=-e^{\phi} \, \frac{1}{2} (*F_4)^{mn} \nn \\
\tilde \cf_{22}&=e^{\phi} \, {*F_{6}} \ , \qquad
 \tilde \cf_{mn2}^{\,\,\,\quad1}=-e^{\phi} \, \frac{1}{2}F_{mn}  \ .
\end{align}

In applying the twisted derivative to the algebraic structures $L$ and $K$, the following tensor products appear
\begin{align}
{\cal D} L= \quad D L\ \, \ \ \ &+ \ \ \  {\cal F} \, L \ , \qquad \qquad  {\cal D} K=  \quad D K\ \, \ \ \ + \ \ \  {\cal F} \, K \nn  \\
\rep{56} \times \rep{56} \ &+ \ \rep{912} \times \rep{56} \qquad \qquad \qquad\rep{56} \times \rep{133} \ + \ \rep{912} \times \rep{133} \nn
\end{align}
If we think of the vacua equations as coming from variations of the Killing prepotentials (\ref{Killprep}), out
of these tensor products of representations, the equations should lie in the ${\bf 133}$ representation for
${\cal D} L$, and in the ${\bf 56}$ in ${\cal D} K$. We give in (\ref{Dlambda4bis})-(\ref{DK7})   the full expression for the twisted derivative of an element in the ${\bf 56}$ and an element in the ${\bf 133}$. In section \ref{n1proof} we rewrite the only components that are non-zero
in the case of ${\cal N}=1$ vacua, i.e. for $L$ and $K$ whose only non-zero components are those in (\ref{nonzerocomp}).

\end{subsection}
 
\begin{subsection}{Equations for $\mathcal{N}=1$ vacua}\label{n1}

By following the same reasoning that leads from the superpotential to the
equations for $\mathcal{N}=1$ vacua in the GCG case, a set of three equations were conjectured in \cite{GLSW} to be the EGG analogue of (\ref{koerber1})-(\ref{koerber3}). While the spinor component in the $\Tsub$ decomposition of each equation reproduced the GCG equations (\ref{koerber1})-(\ref{koerber3}), other representations
did not work. Here, we show that the conjectured equations do work if we introduce two modifications: first,
instead of using dressed bispinors and
an untwisted derivative, we use undressed bispinors and a twisted derivative, appropriately projected onto the $\rep{912}$.
This gets rid of the non gauge invariant terms arising in the vector parts of the equations conjectured in \cite{GLSW}. Second, we add a right hand side to the equations with a single internal index, proportional to the derivative of the warp factor or the dilaton. 

The equations are written in terms of $L$ and $K_a$ using the following parameterisation for the spinors 
\begin{equation} \label{N=1thetas}
\theta^1=\left(\begin{array}{c} a\eta_{+}^{1}\\0\end{array}\right), \qquad \theta^2=\left(\begin{array}{c} 0\\ \bar b\eta_{-}^{2}\end{array}\right)
\end{equation}
With this parameterisation, the combinations that are relevant for ${\mathcal N}=1$ supersymmetry are 
\begin{align}
L'&\equiv e^{2A} L \ , \qquad  \nn \\
\ K'_1&\equiv  e^A r^a K_a=e^A K_1 \ ,  \\
K'_+&\equiv  e^{3A} z^a K_a=e^{3A} (K_3+iK_2) \ . \nn 
\end{align}

In the language of EGG, $\mathcal{N}=1$ supersymmetry requires for $L'$, 
\beq
\cd L' \big|_{\bf{133}}=0 \label{N=1EGG1} \  ,
\eeq
for ${\cal D}{K}'_1|_{\bf 56}$\footnote{We are using the notation in (\ref{slf}), where a tilde denotes the component in the ${\bf 28}'$ representation} 
\begin{align}
(\cd K'_{1})^{mn}=0,  \qquad \qquad \widetilde{(\cd K'_{1})}_{mn}&=0  \ , \nn \\
\quad( \cd K'_{1})^{12} =0, \qquad  \qquad \widetilde{(\cd K'_{1})}_{12}&=0 \ , \label{N=1EGG2}\\
(\cd K'_{1})^{m2}=0, \qquad \qquad  \widetilde{(\cd K'_{1})}_{m1}&=0 \ \nn , 
\end{align}
and for ${\cal D}{ K}'_+|_{\bf 56}$
\begin{align}
(\cd K'_{+})_{mn} - i \widetilde{(\cd K'_{+})}_{mn}&=0  \ , \nn \\
\quad (\cd K'_{+})_{12} - i \widetilde{(\cd K'_{+})}_{12}&=0 \ , \label{N=1EGG3}\\
(\cd K'_{+})^{m2}&=0\ . \nn 
\end{align}
The remaining  components of ${\cal D} K$ (all with one internal index) are proportional to derivatives of the dilaton and warp factor as follows 
\begin{align} 
&( \cd K'_{1})^{m1} =4e^{-2A} \partial_pA K'_+{}^{mp}, \quad \  \widetilde{(\cd K'_{1})}_{m2}=-4 e^{-2A} \partial_p A \, (2K'_{+}{}^{p}{}_{m12}+i \delta^p_m K'_+{}^1{}_2) \label{vectorDK1}  ,\\
&(\cd (e^{-\phi} K'_{+}))^{m1} = -4i e^{-\phi} g^{mp} \partial_p A  K'_+{}^1{}_2 \ , \  \widetilde{(\cd (e^{2A-\phi} K'_{+}))}_{m2}= - e^{2A-\phi} H_{mpq}K'_+{}^{12pq}  \label{vectorDK} \\
& \widetilde{(\cd (e^{-4A+\phi} K'_{+}))}_{m1}= 0  \ . \nn  
\end{align}
The equations for $L$, ${K}_3'$ and ${K_+'}$ in (\ref{N=1EGG1})-(\ref{N=1EGG3}) are respectively the EGG version of (\ref{koerber1}), (\ref{koerber2}) and (\ref{koerber3}).
The vectorial equations are a combination of (\ref{koerber1})-(\ref{koerber3}) plus (\ref{norma}). 

\end{subsection}

\begin{subsection}{From SUSY conditions to EGG equations}\label{n1proof}

We will sketch here the proof that $\mathcal{N}=1$ supersymmetry requires (\ref{N=1EGG1})-(\ref{N=1EGG3}) and leave the details, as well the proof of Eqs (\ref{vectorDK1}), (\ref{vectorDK}), to 
Appendix \ref{app:DLDKvssusy}.

Using (\ref{nonzerocomp}) in (\ref{Dlambda4bis})-(\ref{Dlambda7}), we get that the only nontrivial components of Eq.  (\ref{N=1EGG1}) are 
\begin{align}
(\mathcal{D}L')^{1}_{\,\,\,2}&=-e^{\phi} [iF_{0}+(*F_{6})]L'^{12}+\frac{e^{\phi}}{2} [F_{mn}+i(*F_{4})_{mn}]L'^{mn}  \ , \label{DL4} \\
(\mathcal{D}L')^{1}{}_{m}&=-\nabla_{m} L'^{12} \label{DL1} \\
(\mathcal{D}L')^{m}{}_{2}&=-\nabla_{p} L'^{mp}+\frac{i}{2}(*H)^{mnp}L'_{np} \label{DL2}\\
(\mathcal{D}L')_{mnp2}&= \frac{3i}{2}\nabla_{[m} L'_{np]}+ \frac{1}{2}H_{mnp} L'^{12} \label{DL6} \ ,
\end{align}
where we used \eqref{pw1}, while for $K_1'$ we get  
 \begin{align}
 (\mathcal{D} K'_1)^{mn}&=-2\nabla_p K'_1{}^{mnp2}+(*H)^{mnp}K'_1{}^{2}{}_{p}\label{aDK1} \\
 \widetilde{(\mathcal{D} K'_1)}_{mn}&=-2\nabla_{[m} K'_1{}^{2}{}_{n]}\label{aDK2}\\
  \widetilde{ (\mathcal{D} K'_1)}_{12}&=-\nabla_nK'_1{}^{n}{}_{1}-\frac{1}{3}H_{npq} K'_1{}^{2npq} \label{aDK3} \\
 (\mathcal{D} K'_1)^{m1}&=e^{\phi} F_{0} K'_1{}^{m}{}_{1}-e^{\phi} (*F_{4})^{mn}K'_1{}^{2}{}_{n}-e^{\phi} F_{np} K'_1{}^{2npm} \label{aDK4} \\
 \widetilde{(\mathcal{D} K'_1)}_{m2}&=-e^{\phi} {*F_{6}} K'_1{}^{2}{}_m-e^{\phi} F_{mn}K'_1{}{}^{n}{}_{1}+e^{\phi} (*F_{4})^{np}K'_{1\,1npm} \label{aDK6}
 \end{align}
 and for ${ K'}_+$
 \begin{align}
 (\mathcal{D} { K}'_+)^{mn}&=-2\nabla_p {K_+'}^{mnp2}+(*H)^{mnp} {K_+'}^{2}{}_{p}+e^{\phi} (*F_{4})^{mn} {K_+'}^{2}{}_{1} \label{DK11} \\
\widetilde{(\mathcal{D} {K_+'})}_{mn}&=-2\nabla_{[m} {K'_+}^{2}{}_{n]}+e^{\phi} F_{mn}{K_+'}^{2}{}_{1} \label{DK4}\\
 (\mathcal{D} {K_+'})^{m1}&=2\nabla_p {K_+'}^{mp12}+e^{\phi} F_{0}{K'_+}^{m}{}_{1}- e^{\phi} (*F_{4})^{mn}{K_+'}^{2}{}_{n}-e^{\phi} F_{np}{K_+'}^{2npm} \label{DK21}\\
\widetilde{(\mathcal{D} {K_+'})}_{m1}&=-\nabla_m{K_+'}^{2}{}_{1} \label{DK51}\\
  \widetilde{(\mathcal{D} {K_+'})}_{m2}&=-\nabla_p{K_+'}^{p}{}_{m}- H_{mpq} {K_+'}^{pq12}-e^{\phi} {*F_{6}} {K_+'}^{2}{}_{m}-e^{\phi} F_{mp}{K_+'}^{p}{}_{1} \nn \\
 & \quad + e^{\phi} (*F_{4})^{pq}  {K'}_{+\,1pqm} \label{DK61}\\
  (\mathcal{D} {K_+'})^{12}&=-e^{\phi} F_{0} {K'_+}^{2}{}_{1} \label{dk12z}\\
\widetilde{(\mathcal{D} {K_+'})}_{12}&=-\nabla_n{K_+'}^{n}{}_{1}-\frac{1}{3}H_{npq}{K_+'}^{2npq} -e^{\phi} {*F_{6}} {K'_+}^{2}{}_{1} \label{DK71} 
 \end{align}
where we should keep in mind that the components of ${ K}_+$ with an odd (even) number of internal indices are proportional to $K_2$ ($K_3$) (see \eqref{nonzerocomp}). 

We now show that supersymmetry requires (\ref{N=1EGG1}), in particular the components appearing in (\ref{DL4}) and (\ref{DL1}). The proof for the rest 
of the components is in Appendix \ref{app:DlambdaN=1}. 

It is not hard to show that exactly the same combination of RR fluxes appearing on the right hand side of (\ref{DL4}) 
is obtained by multiplying Eq. (\ref{extgravitinopiN=1}), coming from the external gravitino variation,
 by $\Gamma^2$, and tracing over the spinor indices, namely
\begin{equation}
0=\sqrt{2} \, \Tr \left(i\Gamma^2 \Deltamu \pi'\right)=-e^{\phi} [iF_{0}+(*F_6) ] L'^{12}+\frac{e^{\phi}}{2}\left[F_{mn}+i(*F_4)_{mn}\right] L'^{mn} =({\cal D}L')^1{}_2\label{combk1} \nn
\end{equation}
 where in the second equality the term proportional to the derivative of the warp factor goes away by symmetry, and we have used (\ref{pw1}) to relate the $SU(8)$ and $SL(8)$ components of $L$. Supersymmetry requires therefore $({\cal D}L')^1{}_2=0$.

For the equations that involve a covariant derivative of $L^{ab}$, we
use (\ref{intgravitinolambdaappN=1}) coming from
the internal gravitino variation, multiplied by $\Gamma^{ab}$ and we trace over the spinor indices (see Eq. (\ref{pw1})).  
For $ab=12$, for example, this gives
\begin{equation} \label{covderlambda12}
0=\tfrac{\sqrt2}{4} \Tr\left(\Gamma^{12}\Deltam L' \right)=\nabla_mL'^{12} -\partial_m (2A-\phi) L'^{12}- \frac{i}{4}H_{mnp} L'^{np}+\frac{e^{\phi}}{8} [F_{pq}+i(*F_4)_{pq}]\pi'^{2pq}{}_{m} \nn
\end{equation}
where $\pi'$ is defined in (\ref{pi}) and (\ref{piabcd}).
Now we use Eqs. \eqref{extgravitinolambdaN=1} and \eqref{dilatinolambdaN=1} multiplied by $\Gamma_m$ and traced over the spinor indices to cancel the terms containing derivatives of the dilaton and warp factor. In doing this, the term involving $H$ and $F$ fluxes  completely cancel, i.e.
\begin{align}
0&=\frac{\sqrt 2}4 \Tr\left(\Gamma^{12} \Deltam L' +i \Gamma_m (-2\Deltamu L'+\Deltaphi L')\right)\nn\\
&=\nabla_m L'^{12} \nn \\
&=({\cal D}L')^1{}_m \nn \ .
\end{align}

We show in Appendix \ref{app:DlambdaN=1} how supersymmetry requires the remaining equations, \eqref{DL2} and \eqref{DL6}, to vanish.

The equations for $K$ work similarly. For example, to show that (\ref{aDK1}) should vanish, we use (\ref{iu1}) coming from internal gravitino, in the following way  
\begin{align}
0=& -\frac{i}{4}\Tr\left[\Gamma^{mnp2} (e^A \Deltap K_1)\right] & \nn \\
=&-2 e^{A-\phi} \nabla_p (e^{\phi} K_1{}^{mnp2}) 
+\frac{1}{2}H^{mnp} K'_1{}^{1}{}_{p}+\frac{3}{2}(*H)^{mnp} K'_1{}^{2}{}_{p} \nn \\
&- 2e^{-2A+\phi}F_0 K_+'^{mn12}- e^{-2A+\phi}F^{[m|p}K_+{}_{p}{}^{|n]}  \label{dkmu} \ .
\end{align}
We combine this with external gravitino equations (\ref{egra1}), (\ref{egra2}) and dilatino equations (\ref{dila1}), (\ref{dila2}) to get (see more details in Appendix \ref{app:DKN=1}) 
\begin{align}
0=& -\frac{i}{4} \, \Tr \left[\Gamma^{mnp2} (e^A \Deltap K_1)+ \{\Gamma^{mn1},\Deltamu K'_1-\Deltaphi K'_1\}\right]\nn\\
=&-2 \nabla_p K'_1{}^{mnp2}+(*H)^{mnp} K'_1{}^{2}{}_p  \nn \\
=& (\mathcal{D} K'_1)^{mn}
\end{align}
where we have used the notation in (\ref{commut}).

We give the details about the rest of the components of  the twisted derivative of $K'_1$ and $K'_+$ in Appendix \ref{app:DKN=1}.

We will now connect the equations found to their generalized complex geometric counterparts, Eqs. (\ref{koerber1})-(\ref{koerber3}) and (\ref{norma}). Eqs. (\ref{DL1})-(\ref{DL6})) reduce to (\ref{koerber1}). The right hand side of Eq. (\ref{DL4}) is proportional to $\langle F, \Phi^{+}\rangle$, which
can be seen to vanish by wedging  (\ref{koerber1}) with $C^-$ (this means that actually (\ref{DL4}) can be derived from (\ref{DL1})-(\ref{DL6})). 
The  $mn$ and $12$ components of the EGG equations for $K_1'$ and $K_+'$ combine to build up respectively  (\ref{koerber2}) and (\ref{koerber3}). Interestingly, Eq. (\ref{norma}), which is not part of the pure spinor equations but has to be added 
by hand in the GCG language, becomes one of the EGG equations, namely the one on the second line of (\ref{vectorDK}). This can be seen by using (\ref{DK51}) and the fact that $K_+'{}^2{}_1=K_3'{}^2{}_1=-\tfrac{i}{4} e^{3A-\phi} (|\eta_1|^2+|\eta_2|^2)$ 
The other vectorial components of ${\cal D} K$ involve for example terms of the form $\langle F, \Gamma^A \Phi^{-}\rangle$, which making use of (\ref{koerber1})-(\ref{koerber3}), can be shown to be proportional to derivatives of the warp factor. 

Since (\ref{koerber1})-(\ref{norma}) were shown in \cite{GMPT3} to be equivalent to supersymmetry conditions, we conclude 
that the EGG equations (\ref{N=1EGG1})-(\ref{vectorDK}) are completely equivalent to requiring $\mathcal{N}=1$ supersymmetry, i.e., supersymmetry requires (\ref{N=1EGG1})-(\ref{vectorDK}), and (\ref{N=1EGG1})-(\ref{vectorDK}) implies supersymmetry.

As mentioned in section \ref{lk}, $L$ defines an $\Ex6$ structure in $\E7$. We have shown here that  $\mathcal{N}=1$ supersymmetry 
requires this structure to be twisted closed, upon projection to the {\bf 133}. It would be very nice to show that this is equivalent  to the structue being integrable\footnote{Unlike the case of generalized complex structures, even if there is an exceptional Courant bracket \cite{PW},   there is no known correspondence between the differential conditions on the structure and closure of a subset (defined by the structure) of the exceptional generalized tangent bundle under the exceptional Courant bracket.}. For constant warp factor and dilaton, also $K'_1$ is twisted closed. Most of the components of $K'_+$ are also twisted closed after projection onto holmorphic indices in the ${\bf 56}$. The vectorial components of ${\cal D} K$ are proportional to derivatives of the warp factor and dilaton, except 
the second equation in (\ref{vectorDK}), which does not seem to be expressible in terms of such derivatives. 

\end{subsection}

\end{section}


\subsection*{Acknowledgements}
We would like to thank Diego Marqu\'es, Hagen Triendl and especially Daniel Waldram for many useful discussions. 
This work is supported by the DSM CEA-Saclay and by the ERC Starting Independent Researcher Grant 259133 -- ObservableString.

\appendix

\section{$E_{7(7)}$ basics and tensor products of representations}\label{E7}

$E_{7(7)}$ can be defined as the subgroup of $Sp(56,\mathbb{R})$ which in addition to preserve the symplectic structure $\mathcal{S}(\lambda,\lambda')$, preserves also a totally symmetric quartic invariant.
We exploit the decomposition of $E_{7(7)}$ representations under two subgroups \begin{enumerate}
\item{$SL(2,\mathbb{R})\times O(6,6)$ is the physical subgroup appearing as the factorization of (``heterotic") S-duality and the T-duality group that emerges in the framework of generalized geometry}
\item{$SL(8,\mathbb{R})$. This subgroup contains the product $SL(2,\mathbb{R})\times GL(6,\mathbb{R})$, and allows to make contact with $SU(8)/ \mathbb{Z}_2$, the maximal compact subgroup of $E_{7(7)}$. The latter is the group under which the spinors transform, and therefore the natural language to formulate supersymmetry via the Killing spinor equations.}
\end{enumerate}

\subsection{$SL(2,\mathbb{R})\times O(6,6)$}
The fundamental \textbf{56} representation decomposes as
\begin{align}
\nu&=(\nu^{iA},\nu^{+}) \nn \\
\bf{56}&=\bf{(2,12)}+\bf{(1,32)} \nn
\end{align}
For the adjoint \textbf{133} of $\E7$ we have
\begin{align}
\mu&=(\mu^{i}_{\,\,\,j},\mu^{A}_{\,\,\,B},\mu^{i-}) \nn \\
\bf{133}&=\bf{(3,1)}+\bf{(1,66)}+\bf{(2,32')} \nn
\end{align}
where $\mu^i{}_i=0$ and $\mu^{AB}=\mu^A{}_C \, \eta^{CB}$ is antisymmetric.
The \textbf{912} decomposes as
\begin{align}
\phi&=(\phi^{iA},\phi^{i}{}_{j}{}^+,\phi^{A-},\phi^{iABC}) \nn \\
\bf{912}&=\bf{(2,12)}+\bf{(3,32)}+\bf{(1,352)}+\bf{(2,220)} \nn
\end{align}
where $\Gamma_A \Phi^{A-}=0$ and $\phi^{iABC}$ is fully antisymmetric in $ABC$. 

There are various tensor products projected on some particular representation that are used throughout the paper. These are: \\\\
$\bf{56}\times\bf{56}\big|_{\bf{1}}$ (i.e. the symplectic invariant)
\beq \label{symplTsub}	
{\cal S}(\nu, \hat \nu)=\epsilon_{ij} \eta_{AB} \nu^{iA} \hat \nu^{jB} + \langle \nu^+, \hat \nu^+ \rangle
\eeq
$\bf{56}\times\bf{56}\big|_{\bf{133}}$
\begin{align}
(\nu\cdot\hat \nu)^{i}_{\,\,\,j}&=2\epsilon_{jk}\eta_{AB} \nu^{iA} \hat \nu^{kB} \nn \\
(\nu\cdot\hat \nu)^{A}_{\,\,\,B}&=2\epsilon_{ij}(\nu^{iA}\hat \nu^{j}{}_{B}+\hat \nu^{iA}\nu^{j}{}_{B})+\langle \nu^{+},\Gamma^{A}_{\,\,\,B}\hat \nu^{+}\rangle \label{pd1} \\
(\nu\cdot\hat \nu)^{i-}&=\nu^{iA}\Gamma_A \hat \nu^{+}+\hat \nu^{iA}\Gamma_A\nu^{+} \nn  ; 
\end{align}
$\bf{56}\times\bf{133}\big|_{\bf{56}}$
\begin{align}
(\nu \cdot \mu)^{iA}&=\mu^{i}_{\,\,\,j}\nu^{jA}+\mu^{A}_{\,\,\,B}\nu^{iB}+\langle \mu^{i-},\Gamma^{A}\nu^{+}\rangle \nn \\
(\nu \cdot \mu)^{+}&=\frac{1}{4}\mu_{AB}\Gamma^{AB}\nu^{+}+\epsilon_{ij}\nu^{iA}\Gamma_A\mu^{j-} \ ; \label{pd2}
\end{align}
the adjoint action on the adjoint, i.e. $\bf{133}\times\bf{133}\big|_{\bf{133}}$ \ ;
\begin{align} \label{adjac}
(\mu \cdot \hat \mu)^{i}_{\,\,\,j}&=\hat \mu^{i}_{\,\,\,\,k}\mu^{k}_{\,\,\,j}-\mu^{i}_{\,\,\,k}\hat \mu^{k}_{\,\,\,j}+\epsilon_{jk}(\langle\hat \mu^{i-},\mu^{k-}\rangle-\langle \mu^{i-},\hat \mu^{k-}\rangle) \nonumber \\
(\mu \cdot \hat \mu)^{A}_{\,\,\,B}&=\hat \mu^{A}_{\,\,\,C}\mu^{C}_{\,\,\,B}-\mu^{A}_{\,\,\,C}\hat \mu^{C}_{\,\,\,B}+\epsilon_{ij}\langle \hat \mu^{i-},\Gamma^{A}_{\,\,\,B}\mu^{j-}\rangle\\
(\mu \cdot \hat \mu)^{i-}&=\hat \mu^{i}_{\,\,\,j}\mu^{j-}-\mu^{i}_{\,\,\,j}\hat \mu^{j-}+\frac{1}{4}\hat \mu_{\,AB}\Gamma^{AB}\mu^{i-}-\frac{1}{4}\mu_{AB}\Gamma^{AB}\hat \mu^{i-} \nn 
\end{align}
and $\bf{56} \times \bf{133}\big|_{\bf{912}}$ 
\begin{align} \label{56x133=912Tsub}
(\nu \cdot \mu)^{iA}&=\mu^{i}_{\,\,\,j}\nu^{jA}+\mu^{A}_{\,\,\,B}\nu^{iB}+ \langle \nu^+, \Gamma^A \mu^{i-} \rangle \nn \\
(\nu \cdot \mu)^{i}{}_{j}{}^+&=\mu^{i}_{\,\,\,j}\nu^{+}-\epsilon_{jk}\nu^{(i|A}\Gamma_A\mu^{k)-} \\
(\nu \cdot \mu)^{A-}&=-\mu^{A}_{\,\,\,B}\Gamma^{B}\nu^{+}+\frac{1}{10} \mu_{BC} \Gamma^{ABC} \nu^+ +\epsilon_{ij}\nu^{iA}\mu^{j-}-\frac{1}{11} \epsilon_{ij} \nu^{iB} \Gamma_B{}^A \mu^{j-} \nn \\
(\nu \cdot \mu)^{iABC}&=3\nu^{i[A}\mu^{BC]} + \langle \nu^+, \Gamma^{ABC} \mu^{i-} \rangle \nn \ .
\end{align}
\subsection{$SL(8,\mathbb{R})$}
The decomposition of the $\E7$ representations we use in terms of $SL(8,\mathbb{R})$ are the following.\\
For the fundamental \textbf{56} we have
\begin{align}
\nu&=(\nu^{ab},\tilde \nu_{ab}) \nn \\
\bf{56}&=\bf{28}+\bf{28}'\label{fundsl8app} \ .
\end{align}
with $\nu^{ba}=-\nu^{ab}$.\\
The adjoint \textbf{133} decomposes as
\begin{align}
\mu&=(\mu^{a}_{\,\,\,b},\mu_{abcd}) \nn \\
\bf{133}&=\bf{63}+\bf{70} \label{adjsl8app}
\end{align}
with $\mu^a{}_a=0$, and $\mu_{abcd}$ fully antisymmetric.\\
For the \textbf{912} we have
\begin{align}
\phi&=(\phi^{ab},\phi^{abc}{}_{d},\tilde \phi_{ab},\tilde \phi_{abc}{}^{d}) \nn \\
\bf{912}&=\bf{36}+\bf{420}+\bf{36}'+\bf{420}'\label{ntsl8}
\end{align}
with $\phi^{ab}=\phi^{ba}$, $\phi^{abc}{}_d=\phi^{[abc]}{}_d$ and $\phi^{abc}{}_c=0$ and similarly for the tided objects. \\
The $\SLE$ decomposition of the tensor products is the following.\\
The adjoint action on the fundamental, $\bf{56}\times\bf{133}\big|_{\bf{56}}$ is\footnote{Note
tht this convention differs by a sign in the $\star \mu$ term than the one used in \cite{PW,CJ}. This choice is correlated with the 
choice in (\ref{Kasd}), and affects a few signs in the equations that follow (those in the terms involving $\star \mu$).}.
\begin{align} \label{56x133=56SLE}
(\nu \cdot \mu)^{ab}&=\mu^{a}_{\,\,\,c}\nu^{cb}+\mu^{b}_{\,\,\,c}\nu^{ac}+\star\mu^{abcd}\tilde \nu_{cd}\\
(\nu \cdot \mu)_{ab}&=-\mu^{c}_{\,\,\,a}\tilde \nu_{cb}-\mu^{c}_{\,\,\,b}\tilde \nu_{ac}-\mu_{abcd}\nu^{cd} \nn
\end{align} The symplectic invariant $\bf{56}\times\bf{56}\big|_{\bf{1}}$ reads
\beq \label{symplSLE}
{\cal S}(\nu,\hat \nu)= \nu^{ab} \tilde{\hat \nu}_{ab}-\tilde \nu_{ab} \hat \nu^{ab}
\eeq
The $\bf{56}\times\bf{56}\big|_{\bf{133}}$ reads 
\begin{align} \label{56x56=133SLE}
(\nu\cdot \hat \nu)^{a}_{\,\,\,b}&=(\nu^{ca}\tilde{\hat \nu}_{cb}-\frac{1}{8}\delta^{a}_{\,\,\,b}\nu^{cd}\tilde{\hat \nu}_{cd})+(\hat \nu^{ca}\tilde \nu_{cb}-\frac{1}{8}\delta^{a}_{\,\,\,b}\hat \nu^{cd}\tilde \nu_{cd})\\
(\nu\cdot \hat \nu)_{abcd}&=-3(\tilde \nu_{[ab}\tilde{\hat \nu}_{cd]} + \frac{1}{4!} \epsilon_{abcdefgh} \nu^{ef} \hat \nu^{gh}) \nn 
\end{align}
where $\star \mu$ is the 8-dimensional Hodge dual, while the adjoint action on the adjoint $\bf{133}\times\bf{133}\big|_{\bf{133}}$ gives
\begin{align} \label{133x133=133SLE}
(\mu \cdot \hat \mu)^{a}_{\,\,\,b}&=(\mu^{a}_{\,\,\,c}\hat \mu^{c}_{\,\,\,b}-\hat \mu^{a}_{\,\,\,c}\mu^{c}_{\,\,\,b})-\frac{1}{3}(\star \mu^{acde}\hat \mu_{bcde}-\star\hat \mu^{acde}\mu_{bcde})\\
(\mu \cdot \hat \mu)_{abcd}&=4(\mu^{e}_{\,\,\,[a}\hat \mu_{bcd]e}-\hat \mu^{e}_{\,\,\,[a}\mu_{bcd]e}) \nn 
\end{align}
The $\bf{56} \times \bf{133}\big|_{\bf{912}}$ is 
\begin{align} \label{56x133=912SLE}
(\nu \cdot \mu)^{ab}&=(\nu^{ac}\mu^{b}_{\,\,\,c}+\nu^{bc}\mu^{a}_{\,\,\,c}) \nn \\
(\nu \cdot \mu)_{ab}&=- (\tilde \nu_{ac}\mu^{c}_{\,\,\,b}+\tilde \nu_{bc}\mu^{c}_{\,\,\,a}) \nn \\
(\nu \cdot \mu)^{abc}{}_{d}&=-3(\nu^{[ab}\mu^{c]}_{\,\,\,\,b}-\frac{1}{3}\nu^{e[a}\mu^{b}_{\,\,\,e}\delta^{c]}_{\,\,\,d})+2(\tilde \nu_{ed} \star\mu^{abce}+\frac{1}{2}\tilde \nu_{ef}\star\mu^{ef[ab}\delta^{c]}_{\,\,\,d})\\
(\nu \cdot \mu)_{abc}{}^d&=-3(\tilde \nu_{[ab}\mu^{d}_{\,\,\,c]}-\frac{1}{3}\tilde \nu_{e[a}\mu^{e}_{\,\,\,b}\delta^{d}_{\,\,\,c]})+2(\nu^{ed} \mu_{abce}+\frac{1}{2}\nu^{ef}\mu_{ef[ab}\delta^{d}{}_{c]}) \nn 
\end{align}
The $\bf{912}\times\bf{56}\big|_{\bf{133}}$ gives
\begin{align} \label{912x56=133SLE}
(\phi \cdot \nu)^{a}_{\,\,\,b}&=(\nu^{ca}\tilde \phi_{cb}+\tilde \nu_{cb}\phi^{ca})+(\tilde \nu_{cd}\phi^{cda}{}_{b}-\nu^{cd}\tilde \phi_{cdb}{}^a)\\
(\phi \cdot \nu)_{abcd}&=-4( \tilde \phi_{[abc}{}^e\tilde \nu_{d]e}-\frac{1}{4!}\epsilon_{abcdm_1m_2m_3m_4}\phi^{m_1m_2m_3}{}_{e}\nu^{m_4 e}) \nn 
\end{align}
and finally $\bf{912}\times\bf{133}\big|_{\bf{56}}$ is
\begin{align} \label{912x133=56SLE}
(\phi \cdot \mu)^{ab}&=-(\phi^{ac}\mu^{b}_{\,\,\,c}-\phi^{bc} \mu^{a}_{\,\,\,c})-2\phi^{abc}{}_{d}\mu^{d}{}_{c}\nonumber\\&+\frac{2}{3}(\tilde \phi_{m_1m_2m_3}{}^a \star\mu^{m_1m_2m_3 b}-\tilde \phi_{m_1m_2m_3}{}^b \star\mu^{m_1m_2m_3 a})\\
(\phi \cdot \mu)_{ab}&=(\tilde \phi_{ac}\mu^{c}{}_{b}-\tilde \phi_{bc}\mu^{c}_{\,\,\,a})-2 \tilde \phi_{abc}{}^d \mu^{c}{}_{d}\nonumber\\&-\frac{2}{3}(\phi^{m_1m_2m_3}{}_{b}\, \mu_{m_1m_2m_3a}-\phi^{m_1m_2m_3}{}_{a} \, \mu_{m_1m_2m_3b}) \label{signrev}
\end{align}

\section{$SU(8)$ and $SU(4) \times SU(2)$ conventions}\label{gamma}

The spinor $\theta^{\alpha}$ transforms in the fundamental representation of $SU(8)$. The standard interwining relations
\begin{equation}
\Gamma_M^{\dagger}=A\Gamma_M A^{-1},\qquad \Gamma_M^{\,\,\,T}=C^{-1}\Gamma_MC, \qquad (\Gamma_M)^{*}=-D^{-1}\Gamma_MD
\end{equation}
allow to define the conjugate spinors
\beq \label{thetabar}
\bar \theta=\theta^{\dagger} A \ , \quad \theta^t=C \theta^T \ , \quad \theta^c=D \theta^* \ .
\eeq
Under $SU(8)$, the $\rep{56}$ decomposes according to
\begin{align}
\nu&=(\nu^{\alpha\beta},\bar{\nu}_{\alpha\beta})\nonumber\\
\bf{56}&=\bf{28}+\bar{\bf{28}}\label{suf}
\end{align}
while for the adjoint \textbf{133} we have
\begin{align}
\mu&=(\mu^{\alpha}_{\,\,\,\beta},\mu^{\alpha\beta\gamma\delta},\bar{\mu}_{\alpha\beta\gamma\delta})\nonumber\\
\bf{133}&=\bf{63}+\bf{35}+\bar{\bf{35}}\label{sua} \ .
\end{align}
where $\mu^\alpha{}_\alpha=0$ and $\bar{\mu}_{\alpha\beta\gamma\delta}=\star \mu_{\alpha\beta\gamma\delta}$.
Note that these are very similar to the $\SLE$ decompositions (\ref{fundsl8app}), (\ref{adjsl8app}).  To go from one to the other, we use   for the $\rep{56}$ \cite{PW}
\begin{align} 
\nu^{ab}&=\frac{\sqrt 2}{8} (\nu^{\alpha\beta}+ \bar \nu^{\alpha\beta}) \Gamma^{ab}{}_{\beta\alpha} \ , \\
\tilde \nu_{ab}&=-\frac{\sqrt 2}{8}i (\nu^{\alpha\beta}- \bar \nu^{\alpha\beta}) \Gamma^{ab}{}_{\beta\alpha} \ . 
\end{align}
In the main text we use a complex $\rep{28}$ object, defined from its real pieces $\lambda^{ab},\tilde \lambda_{ab}$ in the obvious way
\beq\label{pw1}
L^{ab}=\lambda^{ab}+i \tilde \lambda^{ab}=\frac{\sqrt 2}{4} L^{\alpha\beta} \Gamma^{ab}{}_{\beta\alpha}
\eeq
From the $\rep{63}$ adjoint representation of $SU(8)$ 
(i.e. taking 
 $\mu_{\alpha\beta\gamma\delta}=0$) one recovers the following $\SLE$ components 
\begin{align}
\mu_{ab}&=-\frac{1}{4} \mu^{\alpha}{}_{\beta} \Gamma_{ab}{}^{\beta}{}_{\alpha} \nonumber\\
\mu_{abcd}&= \frac{i}{8} \mu^{\alpha}{}_{\beta} \Gamma_{abcd}{}^{\beta}{}_{\alpha}  \label{pw2}
\end{align}
 where $\mu_{ba}=-\mu_{ab}$ and $\star \mu_{abcd}=-\mu_{abcd}$
(the symmetric and self-dual pieces are obtained from the $\rep{70}$ representation $\mu^{\alpha\beta\gamma\delta}$)
and $\mu_{ab}=g_{ac} \mu^c{}_b$ (at this point there is a metric since $SL(8) \cap SU(8)=SO(8)$).

When breaking $SU(8) \to SU(4) \times SU(2)$, the spinor index decomposes in a pair of indices $\alpha=\hat \alpha I$,
where $\hat \alpha$ is an $SU(4)$ spinor index. For the  \textit{Cliff}$(8,0)$ gamma matrices, we have used the following 
basis in terms of \textit{Cliff}$(6,0)$ and Pauli sigma-matrices
\begin{align} \label{gammabasis}
\Gamma^m{}^{\alpha}_{\,\,\,\beta}&=
\gamma^m \otimes \sigma_3 \nonumber \\
\Gamma^1{}^{\alpha}_{\,\,\,\beta}&=
\mathbb{I}_6 \otimes \sigma_1 \\
\Gamma^2{}^{\alpha}_{\,\,\,\beta}&=
\mathbb{I}_6 \otimes \sigma_2 \nonumber \ . 
\end{align}

The intertwiners $A,C,D$ also split into \textit{Cliff}$(6) \otimes$\textit{Cliff}$(2)$ intertwiners. In particular, $C$ splits as
\begin{equation}
C=\hat{C}\otimes c
\end{equation}
where $\hat C$ is the intertwiner
\begin{equation}
\gamma^{mT}=-\hat{C}^{-1}\gamma^m\hat{C} \ .
\end{equation}
We get that
\begin{equation}
C_{\alpha \beta}=\hat{C}\otimes \sigma_{1}
\end{equation}

We will use a basis for the \textit{Cliff}$(6,0)$ gamma matrices in which $\hat A=\hat C=\hat D=\mathbb{I}$, and therefore
the $SU(4)$ conjugate spinors are just
\beq
\bar \eta=\eta^{\dagger} \ , \quad  \eta^t=\eta^T \ , \quad  \eta^c=\eta^*
\eeq
and $\eta_-\equiv \eta_+^*$. In this basis, the $SU(8)$ spinors in \eqref{fulla} have conjugates
\begin{align}
\theta^{1t}&=(0,\eta^{1T}_{-}) \\
\bar \theta_1=\theta^{1\dagger}&=(\eta^{1\dagger}_{+},0) \ .
\end{align}

\section{$GL(6,\mathbb{R})$ embedding in  $SL(8,\mathbb{R})$}\label{switch}

The $GL(6,\mathbb{R})$ weights of the different $\Tsub$ representations is worked out in \cite{GLSW}. 
It turns out that the two components  of an $\SLR$ doublet have different $GL(6,\mathbb{R})$ weights. 
To find the $GL(6,\mathbb{R})$ weight in the $SL(8,\mathbb{R})$ decomposition, we use that  
$\SLE \supset SL(2,\mathbb{R})\times GL(6,\mathbb{R})\subset \Tsub$, where the common $GL(6,\mathbb{R})$ piece
corresponds to the diffeomorphisms. Decomposing $a=(m,i)$ with $m=1,..,6$ a $GL(6)$ index and $i=1,2$ an $SL(2)$ index, 
the embedding of $SL(2,\mathbb{R})\times GL(6,\mathbb{R})\subset SL(8,\mathbb{R})$ is the following 
\begin{align}
M^a{}_b=&\left( 
\begin{array}{cc} 
(\mbox{det}a)^{-1/4} a^m{}_n & 0 \\
0 & (\mbox{det}a)^{1/4} \left( \begin{array}{cc} 
(\mbox{det}a)^{-1/2} e^{\phi}&0\\
0&(\mbox{det}a)^{1/2} e^{-\phi} \\ 
\end{array} \right) 
\end{array} \right) \nonumber \\
& =\left( \begin{array}{ccc} 
(\mbox{det}a)^{-1/4} a^m{}_{n} &0&0\\
0&(\mbox{det}a)^{-1/4} e^{\phi} &0 \\
0&0& (\mbox{det}a)^{3/4} e^{-\phi}
\end{array}  \right)\label{high}
\end{align}
where $M \in \SLE, a \in GL(6,\mathbb{R})$, and we have added explicit factors of the dilaton that are needed in order to get the right transformation properties of the connection. Since a six-form transforms by a factor $({\rm det} g)^{1/2}$ (or equivalently $1/{\rm det} a$), we can write the 8-dimensional metric as
\beq
\hat g_{ab}=\left( \begin{array}{ccc} 
(\mbox{det}g)^{-1/4} g_{mn} &0&0\\
0&(\mbox{det}g)^{-1/4} e^{-2\phi}&0 \\
0&0& (\mbox{det}g)^{3/4} e^{2\phi}
\end{array}  \right)\label{SLEmetric}
\eeq

The different $\SLE$ components of $\rep{56}$ representation $\nu=(\nu^{ab}, \tilde \nu_{ab})$ transform therefore according to
\begin{equation}
\label{lambdasl8prop}
\begin{aligned}
   \tilde \nu_{mn} &\in \left(\Lambda^6T^*M\right)^{-1/2}\otimes \Lambda^2T^*M \ ,    &\nu^{mn} \in \left(\Lambda^6T^*M\right)^{-1/2}\otimes \Lambda^4T^*M \\
  \tilde \nu_{1m} &\in {\cal L} \otimes  \left(\Lambda^6T^*M\right)^{-1/2}\otimes  T^*M \ ,   & \nu^{1m} \in {\cal L}^{-1} \otimes\left(\Lambda^6T^*M\right)^{-1/2}\otimes  \Lambda^5 T^*M \\
   \tilde \nu_{2m} &\in {\cal L}^{-1} \otimes \left(\Lambda^6T^*M\right)^{-1/2}\otimes (T^*M \otimes \Lambda^6 T^*M) \ ,  & \nu^{2m} \in {\cal L}\otimes\left(\Lambda^6T^*M\right)^{-1/2}\otimes  TM\\
   \tilde \nu_{12} &\in \left(\Lambda^6T^*M\right)^{-1/2}\otimes \Lambda^6 T^*M \ ,  & \nu^{12} \in \left(\Lambda^6T^*M\right)^{-1/2} 
\end{aligned}
\end{equation}
where we have introduced a trivial real line bundle ${\cal L}$ with sections $e^{-\phi} \in {\cal L}$ to account for factors of the dilaton. 
The adjoint  $\mu=(\mu^a{}_b,\mu_{abcd})$ has the following $GL(6,\mathbb{R})$ and dilaton assignments 
\begin{equation}
\label{musl8prop}
\begin{gathered}
  \mu^1{}_1 = -\mu^2{}_2 \in \bbR \ , \qquad
   \mu^1{}_2 \in {\cal L}^{-2} \otimes \Lambda^6T^*M \ , \qquad
   \mu^2{}_1 \in {\cal L}^2 \otimes \Lambda^6TM \ , \qquad 
    \mu^m{}_n \in TM \otimes T^*M \\   
   \mu^1{}_m \in {\cal L}^{-1} \otimes T^*M , \quad
   \mu^2{}_m \in {\cal L}\otimes \Lambda^5 TM \ , \quad
   \mu^m{}_1 \in {\cal L}\otimes TM \ , \quad 
   \mu^m{}_2 \in {\cal L}^{-1} \otimes \Lambda^5 T^*M   \ , \\
   \mu_{mnpq} \in \Lambda^2TM  \ , \quad \mu_{mnp1}= {\cal L}\otimes \Lambda^3TM \ , \quad \mu_{mnp2}={\cal L}^{-1} \otimes \Lambda^3T^*M \ , \quad  \mu_{mn12} \in \Lambda^2T^*M  
\end{gathered}
\end{equation}
Finally, the $\rep{912}$ multiplied by ${\cal L}\otimes \left(\Lambda^6 T^*M\right)^{-1/2}$
(a T-duality invariant factor), transforms as \begin{align} \label{GL6912SLE}
\phi^{11} & \in  {\cal L}^{-1} \otimes \, {\mathbb R} \ ,  \qquad  \qquad \qquad \qquad \quad \phi'_{11} \in {\cal L}^{3} \otimes \Lambda^6TM  \nn\\
 \phi^{12} & \in  \Lambda^6TM \ ,  \qquad \qquad \qquad \qquad \qquad \phi'_{12} \in   {\mathbb R}\nn \\
\phi^{22} & \in   {\cal L}^{3} \otimes (\Lambda^6TM)^2 \ ,  \quad  \ \ \qquad \qquad \quad \phi'_{22} \in {\cal L}^{-1} \otimes   \Lambda^6 T^*M \nn\\
\phi^{mnp}{}_q& \in    \Lambda^3TM \otimes T^*M \ ,  \ \ \qquad \quad \qquad \phi'_{mnp}{}^q \in  \Lambda^3TM \otimes TM \nn\\
\phi^{mnp}{}_1& \in  {\cal L}^2 \otimes  \Lambda^3TM \ ,  \  \qquad \qquad \qquad \   \ \phi'_{mnp}{}^1 \in   \Lambda^3 TM  \\
\phi^{mnp}{}_2& \in   \Lambda^3T^*M \ ,  \qquad \ \ \ \qquad\qquad \qquad \phi'_{mnp}{}^2 \in  {\cal L}^2 \otimes \Lambda^3 TM \otimes \Lambda^6TM \nn\\
\phi^{mn1}{}_2& \in  {\cal L}^{-1} \otimes  \Lambda^4T^*M \ ,  \quad  \quad \qquad \qquad \phi'_{mn1}{}^2 \in  {\cal L}^{3} \otimes  \Lambda^4 TM \otimes \Lambda^6TM \nn\\
\phi^{mn2}{}_1& \in   {\cal L}^{3} \otimes \Lambda^2TM \otimes \Lambda^6TM\ ,   \ \quad \quad \ \   \phi'_{mn2}{}^1 \in {\cal L}^{-1} \otimes  \Lambda^2 T^*M \nn
\end{align}

\section{Computing the twisted derivative}\label{app:derivative}

We show in the following how to obtain the connection from twisting the Levi-Civita covariant derivative \eqref{LC} by the gauge fields $B$, $\tilde{B}$ and $C^-$ in the $\bf{133}$ representation. Using the Hadamard formula we get for any element $A$ in the adjoint
\begin{align*}
e^{-\cA}\nabla e^{\cA}&=\nabla  +\nabla \cA+\frac{1}{2}[\nabla\cA,\cA]+\frac{1}{6}[[\nabla\cA,\cA],\cA]+\dots
\end{align*}
 Using (\ref{form-alg}) we get in the $\Tsub$ decomposition 
\begin{align}
(e^{B} e^{-\tilde B} e^{-C} \nabla e^{C} e^{\tilde B} e^{-B})^{i}_{\,\,\,j}&= \delta^{i}_{\,\,\,j} \nabla+v^iv_j\nabla \tilde{B}+v^iv_j\langle   \nabla C^-, C^- \rangle \ , \nonumber \\
(e^{B} e^{-\tilde B} e^{-C} \nabla e^{C} e^{\tilde B} e^{-B})^{B}_{\,\,\,C}&= \delta^{B}_{\,\,\,C} \nabla -\nabla B^{B}_{\,\,\,C} \ ,\\
(e^{B} e^{-\tilde B} e^{-C} \nabla e^{C} e^{\tilde B} e^{-B})^{i-}&=v^i(e^{B}\nabla C^-) \nonumber \label{connsl2} \ .
\end{align}
We now promote the Levi-Civita connection $\nabla$ to an element carrying a fundamental $\rep{56}$ index, as in (\ref{LC}):
$D^{\cl{A}}=(v^i \nabla^A,0)$ and $\nabla^A=(0,\nabla_m)$.  Finally, we project to the $\rep{912}$ representation 
using the tensor product $\bf{56}\times \bf{133}\big|_{\bf{912}}$ for the subgroup $SL(2,\mathbb{R})\times O(6,6)$ given in (\ref{56x133=912Tsub}).  We recover the simple result
\beq
{\cal F}^{1}{}_{2}{}^+=-F^+ \ , \qquad {\cal F}^1{}_{mnp}=-H_{mnp} \ ,
\label{conneq2}
\eeq
where $F^+=e^B dC^-$, and all the other components are zero.

One can alternatively express the connection in terms of the $SL(8,\mathbb{R})$ subgroup. The derivative $D^{\cl{A}}$ is given in this case by  
\begin{equation}
D_{m2}=-D_{2m}=\nabla_m\label{dersl8} \ ,
\end{equation}
while all other components are zero. Applying this to the gauge fields in (\ref{calAsl8}), and projecting onto the $\rep{912}$ using (\ref{56x133=912SLE}), we find the following non-vanishing components  
\begin{equation}
\cf^{mnp}{}_{2}=- \frac{1}{2}(* H)^{mnp} \ , \quad
\cf^{mn1}{}_{2} =-\frac{e^{\phi}}{2}(* F_{4})^{mn}  \ , \quad
\tilde \cf_{mn2}{}^{1}=- \frac{e^{\phi}}{2}F_{mp} \ , \quad 
\tilde \cf_{22}=e^{\phi} {*F_{6}} \label{connsl2} \ .
\end{equation}
Notice that the mass parameter $F_{(0)}$ cannot be obtained this way, and should be added by hand. Using (\ref{GL6912SLE}), we note that the component $\phi^{11}$ transforms as a scalar, and we therefore assign 
\begin{equation}
\cf^{11} = e^{\phi} F_0 \ .
\end{equation}

\section{Twisted derivative of $L$ and $K$}

Inserting the $\SLE$ decomposition of the derivative and of the fluxes given respectively in (\ref{LCsl8}) and (\ref{sl8F}), and the corresponding $\SLE$ components of the tensor products given in 
(\ref{56x56=133SLE}) and (\ref{912x56=133SLE}), we get the following expressions for the twisted derivative of 
$\lambda=(\lambda^{ab},\tilde \lambda_{ab})$, projected onto the ${\bf 133}$
\begin{align}
(\mathcal{D}\lambda)^{1}_{\,\,\,1}&=-\frac{1}{4}\nabla_{p}\lambda^{p2}\label{Dlambda4bis}\\
(\mathcal{D}\lambda)^{2}_{\,\,\,2}&=\frac{3}{4}\nabla_{m}\lambda^{m2} \label{Dlambda5}\\
(\mathcal{D}\lambda)^{1}_{\,\,\,2}&=-\nabla_{m}\lambda^{1m}-e^{\phi} (*F_{6}) \lambda^{12}+e^{\phi} F_{0} \tilde \lambda_{12}
+\frac{e^{\phi}}{2}F_{mn}\lambda^{mn}  - \frac{e^{\phi}}{2}(*F_4)^{np}\tilde\lambda_{np} \label{Dlambda4} \\
(\mathcal{D}\lambda)^{m}{}_{2}&=-\nabla_{p}\lambda^{mp}-\frac{1}{2}(*H)^{mnp}\tilde  \lambda_{np} -e^{\phi} (*F_{6}) \lambda^{m2}-e^{\phi} (*F_4)^{mn} \tilde  \lambda_{n1} \label{Dlambda2}\\
(\mathcal{D}\lambda)^{1}_{\,\,\,m}&=\nabla_{m}\lambda^{12}+e^{\phi} F_{0} \tilde \lambda_{1m}+e^{\phi} F_{mn} \lambda^{n2} \label{Dlambda3} \\
(\mathcal{D}\lambda)^{n}_{\,\,\,m}&= \nabla_{m}\lambda^{n2} -\frac{1}{4} g^n{}_m \nabla_{p}\lambda^{p2} \label{Dlambda1}\\
(\mathcal{D}\lambda)_{mnp2}&= -\frac{3}{2}\nabla_{[m}\tilde \lambda_{np]}+\frac{1}{2}H_{mnp}\lambda^{12}-\frac{3}{2}e^{\phi} F_{[mn|} \tilde \lambda_{|p]1}-\frac{e^{\phi}}{2}F_{mnpq} \lambda^{2q} \label{Dlambda6}\\
(\mathcal{D}\lambda)_{mn12}&=-\nabla_{[m}\tilde  \lambda_{n]1}+\frac{1}{2}H_{mnp}\lambda^{p2}  \ . \label{Dlambda7}
\end{align}

To get the twisted derivative of $K$ projected on the ${\bf 56}$, we use the tensor products (\ref{56x133=56SLE}) and (\ref{912x133=56SLE}). We find
  \begin{align}
 (\mathcal{D} K)^{mn}&=-2\nabla_p K^{mnp2}+(*H)^{mnp} K^{2}{}_{p}+e^{\phi} (*F_{4})^{mn} K^{2}{}_{1} \label{DK1} \\
\widetilde{(\mathcal{D} K)}_{mn}&=-2\nabla_{[m} K^{2}{}_{n]}+e^{\phi} F_{mn}K^{2}{}_{1} \label{DK4}\\
 (\mathcal{D} K)^{m1}&=2\nabla_p K^{mp12}+e^{\phi} F_{0}K^{m}{}_{1}- e^{\phi} (*F_{4})^{mn}K^{2}{}_{n}-e^{\phi} F_{np}K^{2npm} \label{DK2}\\
\widetilde{(\mathcal{D} K)}_{m1}&=-\nabla_mK^{2}{}_{1} \label{DK5}\\
 (\mathcal{D} K)^{m2}&=0 \label{DK8} \\
 \widetilde{(\mathcal{D} K)}_{m2}&=-\nabla_pK^{p}{}_{m}- H_{mpq} K^{pq12}-e^{\phi} (*F_{6})K^{2}{}_{m}-e^{\phi} F_{mp}K^{p}{}_{1} \nn \\
 & \quad + e^{\phi} (*F_{4})^{pq}  K_{1pqm} \label{DK6}\\
  (\mathcal{D} K)^{12}&=-e^{\phi} F_{0} K^{2}{}_{1} \label{DK3}\\
\widetilde{(\mathcal{D} K)}_{12}&=-\nabla_nK^{n}{}_{1}-\frac{1}{3}H_{npq}K^{2npq} -e^{\phi} (*F_{6})K^{2}{}_{1} \label{DK7} 
 \end{align}
where we have used that
\begin{equation} \label{Kasd}
\star K^{abcd}=-K^{abcd}
\end{equation}
which is a consequence of fact that  $K$ is purely in the $\rep{63}$ of $SU(8)$.

\section{Supersymmetric variations for the $\mathcal{N}=1$ spinor anstaz}\label{susy}

The supersymmetry transformations of the fermionic fields of type IIA read, in the democratic formulation \cite{demo} 
\begin{equation} 
\delta \psi_M = \nabla_M \epsilon +\frac{1}{4} \sla{H_M} {\cal P} \epsilon + \frac{1}{16} e^{\phi}  
\sum_n  \sla \! {F^{(10)}_{n}} \, \Gamma_{M} {\mathcal P}_n \, \epsilon  \, , \label{svariations}
\end{equation} 
\begin{equation}  
\delta \lambda = \left(\sla{\partial} \phi + \frac{1}{2} \sla \! H {\mathcal P}\right) \epsilon 
+ \frac{1}{8} e^{\phi}  
\sum_n  (5-n)  \sla \! {F^{(10)}_{n}} \,  
{\cal P}_n  \epsilon  \, .\label{svariations2}
\end{equation} 
where ${\cal P}=-\sigma^3$ and ${\cal P}_n= (-\sigma_3)^{n/2} \sigma_1$.
We use the standard decomposition of ten-dimensional gamma matrices
\beq
\gamma^{(10)}_{\mu}= \gamma_{\mu} \otimes 1 \, , \ \ \gamma^{(10)}_m=\gamma_5
\otimes \gamma_m \, ,
\eeq
the Poincare invariant ansatz for the RR fluxes
\beq 
\label{eq:rrfs} 
F^{(10)}_{2n} = F_{2n} + {\rm vol}_4 \wedge {\tilde F}_{2n-4} \, \quad {\rm where}\  {\tilde F}_{2n-4} = 
(-1)^{Int[n]}   *_6 {F}_{10-2n}
\eeq 
and we notice that, according to (\ref{gammabasis}), ${\cal P}=i \Gamma^{12}$, ${\cal P}_0={\cal P}_4=\Gamma^1$, ${\cal P}_2={\cal P}_6=-i \Gamma^2 $, $\gamma^m {\cal P}_0=-i \Gamma^{2m}$ and  
 $\gamma^m  {\cal P}_2=\Gamma^{1m}$,\footnote{\label{foot:nog}To avoid clustering of determinant factors, in this section we use the basis for Cliff(8) gamma matrices in (\ref{gammabasis}) without the determinant factors.}.
 
We use the $\mathcal{N}=1$ spinor ansatz (\ref{thetaN=1}), parameterised using two internal spinors, namely $\theta=\theta^1+\theta^2$, where $\theta^1$, $\theta^2$ given in (\ref{N=1thetas}),  we get from the internal components of the gravitino variation that $\mathcal{N}=1$ supersymmetry requires
\begin{equation} \label{intgravitinoN=1}
\delta \psi_m=0 \  \Leftrightarrow \  \nabla_m \theta^1+\frac{i}{8}H_{mnp}\Gamma^{np12}\theta^1-\frac{e^{\phi}}{8}\sla{\FI}\Gamma_m\theta^2 =0\  ,
\end{equation}
and the same exchaging $1\leftrightarrow 2$, where we have defined
\begin{equation} \label{FI}
\sla{F_i}=-i {\sla}{F_h} \Gamma^2+{\sla}{F_a} \Gamma^1
\end{equation}
in terms of the ``hermitean" and ``antihermitean" pieces of $F$, namely
\beq
F_h = \frac12(F+s(F))=F_{0}+F_{4} \ , \quad F_a =\frac12(F-s(F))= F_{2}+F_{6}
\eeq
and finally
\beq
{\sla}{F}_{(n)}=\frac{1}{n!} F_{i_1...i_n} \Gamma^{i_1...i_n} \ . 
\eeq
We will also need the equations involving $\bar \theta$, which is
\begin{equation}
\nabla_m\bar \theta^{1}-\frac{i}{8}H_{mnp} \bar \theta^1 \Gamma^{np12} +\frac{e^{\phi}}{8} \bar \theta^2 \Gamma_m \sla{\FI} =0 \ ,\label{intgravitinobarN=1}
\end{equation}
From the external gravitino variation, we get that $\mathcal{N}=1$ vacua should satisfy
\begin{equation} \label{extgravitinoN=1}
\delta \psi_\mu=0 \  \Leftrightarrow \ i \sla{\PE} A \, \theta^1 +\frac{e^{\phi}}{4}\sla{\FE}\theta^{2}=0, \quad
\end{equation}
and similarly exchanging 1 and 2, where
\beq \label{FE}
\sla{\FE}= \sla{F_h} \Gamma^1-i {\sla}{F_{a}} \Gamma^{2}
\eeq
and 
\beq
\sla{\PE} A =\partial_m A \, \Gamma^{m12} \ .
\eeq
The hermitean conjugate equation reads
\begin{equation} \label{extgravitinobarN=1}
 i \, \bar \theta^1 \sla{\PE} A \,  +\frac{e^{\phi}}{4} \bar \theta^2 \sla{\FE}=0, \quad
\end{equation}
From the dilatino variation, we get
\begin{equation} \label{dilatinoN=1}
\delta \lambda=0 \  \Leftrightarrow   i \sla{\PE} \phi \, \theta^1 +\frac{1}{12}H_{mnp}\Gamma^{mnp}\theta^1+\frac{e^{\phi}}{4}\sla{\FD}\theta^2=0
\end{equation}
where we have defined
\beq \label{FD}
\sla{\FD}=(5-n) \sla{\FE}  \ . 
\eeq
The hermitean conjugate equation reads
\begin{equation} \label{dilatinobarN=1}
  i \bar \theta^1 \sla{\PE} \phi  -\frac{1}{12}H_{mnp} \bar \theta^1 \Gamma^{mnp}+\frac{e^{\phi}}{4} \bar \theta^2 \sla{\FD}=0
\end{equation}

\section{${\cal D}L$ and ${\cal D} K$ versus $\mathcal{N}=1$ supersymmetry} \label{app:DLDKvssusy}

\subsection{ ${\cal D}L$} \label{app:DlambdaN=1}

Multiplying Eq. (\ref{intgravitinoN=1}) (coming from the internal gravitino variation) for the covariant derivative of $\theta^1$ ($\theta^2$), on the right by $e^{2A-\phi} \theta^2$ ($e^{2A-\phi} \theta^1$), and substracting the two equations, we get the following equation for 
the covariant derivative of $L'$
\begin{equation} \label{intgravitinolambdaappN=1}
(\Deltam L')^{\alpha\beta}\equiv \nabla_{m} L'^{\alpha\beta} -\partial_m (2A-\phi) L'^{\alpha\beta}+ \frac14 (i H_{mnp}\Gamma^{np12} L')^{\alpha \beta}-\frac{e^{\phi}}{4}(\sla{\FI}\Gamma_m \pi')^{\alpha\beta}=0 \ .
\end{equation}
where we have defined
\begin{equation} \label{pi}
\pi'^{\alpha\beta} \equiv e^{2A-\phi} ( \theta^{2}\theta^{2}-\theta^{1}\theta^{1})^{\alpha\beta} \equiv e^{2A-\phi} \pi^{\alpha\beta} \ .
\end{equation} 
We will also need the $SL(8)$ object $\pi^{abcd}$, which we define to be
\beq \label{piabcd}
\pi'^{abcd}=\frac{\sqrt2}{4} \pi'^{\alpha\beta} \Gamma^{abcd}{}_{\beta \alpha}
\eeq

Multiplying (\ref{extgravitinoN=1}) (coming from external gravitino variation on $\theta^1$) by $\theta^2$, and substracting to the equation with $\theta^1$ and $\theta^2$ exchanged, we get the following equation 
\beq \label{extgravitinolambdaN=1}
(\Deltamu L)^{\alpha\beta} \equiv i \partial_m A \, (\Gamma^{m12}L)^{\alpha\beta} +\frac{e^{\phi}}{4}(\sla{\FE} \pi)^{\alpha\beta}=0 \ .
\eeq
If instead we multiply (\ref{extgravitinoN=1}) by $\theta^1$ and substract the corresponding equation for $\theta^2$ multiplied by $\theta^2$, we get 
\beq \label{extgravitinopiN=1}
(\Deltamu \pi)^{\alpha\beta} \equiv i \partial_m A \, (\Gamma^{m12}\pi)^{\alpha\beta} +\frac{e^{\phi}}{4}(\sla{\FE} L)^{\alpha\beta}=0 \ .
\eeq
Doing the same on the dilatino (\ref{dilatinoN=1}) we get
\beq \label{dilatinolambdaN=1}
 (\Deltaphi L)^{\alpha\beta}\equiv i \partial_m \phi \, (\Gamma^{m12}L)^{\alpha\beta}  +\frac{1}{12} H_{mnp}(\Gamma^{mnp} L)^{\alpha\beta}
+\frac{e^{\phi}}{4}(\sla{\FD} \pi)^{\alpha\beta}=0 \ ,
\eeq
and a similar equations with $L$ and $\pi$ exchanged, that will not be used.

We show here how supersymmetry requires equations \eqref{DL1}-\eqref{DL6} to vanish. 
For each of them, we use (\ref{intgravitinolambdaappN=1}) plus $l_e$ times (\ref{extgravitinolambdaN=1}) and $l_d$ times (\ref{dilatinolambdaN=1}), and take in the one to last step 
\beq
 l_e=-2 \ , \quad l_d=1 \ .
 \eeq
We show that susy requires Eq. \eqref{DL1} to vanish by
 \begin{align}
0=&\frac{\sqrt2}{4}\Tr\left(\Gamma^{12} \Deltam L'+i \Gamma_m(l_e\Deltamu+l_d\Deltaphi)L'\right)\nn\\
=&\nabla_m L'^{12} -\partial_m (2A-\phi)L'^{12}-\partial_m (l_eA+ l_d\phi) L'^{12}+\frac{i}{4}(-1+l_d)H_{mpq}L'^{pq}\nn\\
&-\frac{e^{\phi}}{8} [F_{pq}(-1+l_e+3l_d)-i(*F_4)(1+l_e+l_d)]\pi'^{2pq}{}_{m} \nn \\
=&\nabla_m L'^{12} \qquad \quad  \nn \\
=&(\mathcal{D}L')^{1}{}_{m} \ ,
\end{align}
To get \eqref{DL2} we do \begin{align}
0=&\frac{\sqrt{2}}{4}\Tr\left(-\Gamma^{mn}\Deltan L' +i\Gamma^{m12}(l_d\Deltaphi L'+l_e\Deltamu L')\right)\nn\\
=&-\nabla_{p} L'^{mp}+\partial_n( 2A-\phi)L'^{mn}+\partial_{n}(l_eA+l_d\phi)L'^{mn}+\frac{i}{4}( 3-l_d)(*H)^{mpq}L'_{pq}\nn\\
&-\frac{e^{\phi}}{8}[F_{pq}(-1+l_e+3l_d)-i(*F_4)_{pq}(1+l_e+l_d)]\pi'^{1pq}{}_{m} \nn \\
=&-\nabla_{p} L'^{mp}+\frac{i}{2}(*H)^{mnp}L'_{np} \nn \\
=&(\mathcal{D}L')^{m}{}_{2} \nn \ ,
\end{align}
while for \eqref{DL6} we use
\begin{align}
0=&\frac{\sqrt{2}}{8}\Tr\left(3i\Gamma_{[mn|}\Delta_{p]} L'  -\Gamma_{mnp12}(l_d\Deltaphi L'+l_e\Deltamu L')\right)\nn\\
=&\frac{3i}{2}\nabla_{[m}L'_{np]}-\frac{3}{2}i\partial_{[m}(2A-\phi)L'_{|np]}-\frac{3}{2}i\partial_{[m|}(l_eA+l_d\phi)L'_{|np]}\nn\\
&+\frac{1}{4}(3-l_d)H_{mnp}L'_{12}+\frac{3}{4}(-1+l_d)(*H)_{[mn|q}L'^{q}{}_{|p]}\nn\\
&+\frac{e^{\phi}}{8} [F_{0}(-3+l_e+5l_d)-i(*F_6)(3+l_e-l_d)]\pi'_{2mnp}\nn\\
&+3\frac{e^{\phi}}{8}[iF_{[m|q}(-1+l_e+3l_d)+(*F_4)_{[m|q}(1+l_e+l_d)]\pi'^{1q}{}_{|np]} \nn \\
=&\frac{3i}{2}\nabla_{[m} L'_{np]}+ \frac{1}{2}H_{mnp} L'^{12} \nn \\
=&(\mathcal{D}L')_{mnp2} \nn \ .
\end{align}

\newpage{}

\subsection{${\cal D} K$} \label{app:DKN=1}
We define the following quantities
\begin{equation}
K_0'=e^{A}K_0 \ , \qquad K_1'=e^{A}K_1 \ ,  \qquad K_2'=e^{3A}K_2\ ,  \qquad K_3'=e^{3A}K_3 \ . 
\end{equation}
Combining (\ref{intgravitinoN=1}) multiplied by $\bar \theta$ with (\ref{intgravitinobarN=1}) multiplied by $\theta$, we obtain 
\begin{align}
\Deltam K_0&\equiv e^{-\phi}\nabla_m (e^{\phi} K_0){}^{\alpha}{}_{\beta}  +\frac{i}{8}H_{mnp}[\Gamma^{np12}K_0'-K_0'\Gamma^{np12}]^{\alpha}{}_{\beta}-\frac{e^{\phi}}{8}[\sla{F_i}\Gamma_m K_1'-K_1'\Gamma_m\sla{F_i}]^{\alpha}{}_{\beta}=0 \label{iu0} \\
\Deltam K_1&\equiv e^{-\phi} \nabla_m (e^{\phi} K_1){}^{\alpha}{}_{\beta}  +\frac{i}{8}H_{mnp}[\Gamma^{np12}K_1'-K_1'\Gamma^{np12}]^{\alpha}{}_{\beta}-\frac{e^{\phi}}{8}[\sla{F_i}\Gamma_m K_0'-K_0'\Gamma_m\sla{F_i}]^{\alpha}{}_{\beta}=0\label{iu1} \\
\Deltam K_2&\equiv e^{-\phi} \nabla_m (e^{\phi} K_2){}^{\alpha}{}_{\beta}+\frac{i}{8}H_{mnp}[\Gamma^{np12}K_2'-K_2'\Gamma^{np12}]^{\alpha}{}_{\beta}-i\frac{e^{\phi}}{8}[\sla{F_i}\Gamma_mK_3'+K_3'\Gamma_m\sla{F_i}]^{\alpha}{}_{\beta}=0\label{iu2}\\
\Deltam K_3&\equiv e^{-\phi} \nabla_m (e^{\phi} K_3){}^{\alpha}{}_{\beta}+\frac{i}{8}H_{mnp}[\Gamma^{np12}K_3'-K_3'\Gamma^{np12}]^{\alpha}{}_{\beta}+i\frac{e^{\phi}}{8}[\sla{F_i}\Gamma_mK_2'+K_2'\Gamma_m\sla{F_i}]^{\alpha}{}_{\beta}=0 \label{iu3}
\end{align}
where the factors of the dilaton inside the covariant derivatives are there to cancel the explicit dilaton dependence of $K$ (see (\ref{lambdaKsu8})). 

Multiplying the external gavitino or dilatino equation, Eqs. (\ref{extgravitinoN=1}) and (\ref{dilatinoN=1}) by $\bar \theta^2$ on the right, and adding it to the same equation with
$\theta^1$ and $\theta^2$ exchanged, we get 
\begin{align}
(\Deltamu K_1)^{\alpha}{}_{\beta}&\equiv i\partial_m A [\Gamma^{m12}K_1]^{\alpha}{}_{\beta}+\frac{e^{\phi}}{4}[\sla{F_e}K_0]^{\alpha}{}_{\beta}=0\label{egra1} \ , \\
(\Deltaphi K_1)^{\alpha}{}_{\beta}&\equiv i\partial_m \phi [\Gamma^{m12}K_1]^{\alpha}{}_{\beta}+\frac{1}{12}H_{mpq}[\Gamma^{mpq}K_1]^{\alpha}{}_{\beta}+\frac{e^{\phi}}{4}[\sla{F_d}K_0]^{\alpha}{}_{\beta}=0 \label{dila1}\ .
\end{align}

We can also use the complex conjugate equations (\ref{extgravitinobarN=1}), (\ref{dilatinobarN=1}) multiplied on the left by $\theta^2$. This gives 
\begin{align}
(K_1 \Deltamu )^{\alpha}{}_{\beta}&\equiv i\partial_m A [K_1\Gamma^{m12}]^{\alpha}{}_{\beta}+\frac{e^{\phi}}{4}[K_0\sla{F_e}]^{\alpha}{}_{\beta}=0\label{egra2} \ , \\
(K_1 {\Deltaphi} )^{\alpha}{}_{\beta}&\equiv i\partial_m \phi [K_1\Gamma^{m12}]^{\alpha}{}_{\beta}-\frac{1}{12}H_{mpq}[K_1\Gamma^{mpq}]^{\alpha}{}_{\beta}+\frac{e^{\phi}}{4}[K_0\sla{F_d}]^{\alpha}{}_{\beta}=0 \label{dila2}\,\\
\end{align}
We will also need the corresponding equations mixing $K_3$ and $K_2$
\begin{align}
(\Deltamu K_3)^{\alpha}{}_{\beta}\equiv& i\partial_mA [\Gamma^{m12}K_3]^{\alpha}{}_{\beta}-i\frac{e^{\phi}}{4}[\sla{F_e}K_2]^{\alpha}{}_{\beta}=0 \label{extK3l}\\
(K_3 {\Deltamu} )^{\alpha}{}_{\beta}\equiv & i\partial_m A [K_3\Gamma^{m12}]^{\alpha}{}_{\beta}+i\frac{e^{\phi}}{4}[K_2\sla{F_e}]^{\alpha}{}_{\beta}=0 \label{extK3r} \\
(\Deltaphi K_3)^{\alpha}{}_{\beta} \equiv & i\partial_m\phi [\Gamma^{m12}K_3]^{\alpha}{}_{\beta}+\frac{1}{12}H_{mpq}[\Gamma^{mpq}K_3]^{\alpha}{}_{\beta}-i\frac{e^{\phi}}{4}[\sla{F_d}K_2]^{\alpha}{}_{\beta}=0\\
(K_3{\Deltaphi} )^{\alpha}{}_{\beta}\equiv & i\partial_m \phi [K_3\Gamma^{m12}]^{\alpha}{}_{\beta}-\frac{1}{12}H_{mpq}[K_3\Gamma^{mpq}]^{\alpha}{}_{\beta}+i\frac{e^{\phi}}{4}[K_2\sla{F_d}]^{\alpha}{}_{\beta}=0
\end{align}
\begin{align}
(\Deltamu K_2)^{\alpha}{}_{\beta} \equiv & i\partial_mA [\Gamma^{m12}K_2]^{\alpha}{}_{\beta}+i\frac{e^{\phi}}{4}[\sla{F_e}K_3]^{\alpha}{}_{\beta}=0 \label{extK2l}\\
(K_2 {\Deltamu} )^{\alpha}{}_{\beta}\equiv & i\partial_m A [K_2\Gamma^{m12}]^{\alpha}{}_{\beta}-i\frac{e^{\phi}}{4}[K_3\sla{F_e}]^{\alpha}{}_{\beta}=0\\
(\Deltaphi K_2)^{\alpha}{}_{\beta}\equiv&  i\partial_m\phi [\Gamma^{m12}K_1]^{\alpha}{}_{\beta}+\frac{1}{12}H_{mpq}[\Gamma^{mpq}K_1]^{\alpha}{}_{\beta}+i\frac{e^{\phi}}{4}[\sla{F_d}K_3]^{\alpha}{}_{\beta}=0 \label{dilK2l}\\
(K_2 {\Deltaphi} )^{\alpha}{}_{\beta}\equiv & i\partial_m \phi [K_1\Gamma^{m12}]^{\alpha}{}_{\beta}-\frac{1}{12}H_{mpq}[K_1\Gamma^{mpq}]^{\alpha}{}_{\beta}-i\frac{e^{\phi}}{4}[K_3\sla{F_d}]^{\alpha}{}_{\beta}=0
\end{align}
and the following ones involving $K_0$ and $K_1$
\begin{align}
(\Deltamu K_0)^{\alpha}{}_{\beta}&\equiv i\partial_m A [\Gamma^{m12}K_0]^{\alpha}{}_{\beta}+\frac{e^{\phi}}{4}[\sla{F_e}K_1]^{\alpha}{}_{\beta}=0\label{egra3} \ , \\
(K_0 \Deltamu )^{\alpha}{}_{\beta}&\equiv i\partial_m A [K_0\Gamma^{m12}]^{\alpha}{}_{\beta}+\frac{e^{\phi}}{4}[K_1\sla{F_e}]^{\alpha}{}_{\beta}=0\label{egra2} 
\end{align}

Given a generic $K$ and product of gamma matrices $\Gamma^{a_1\dots a_i}$ we will make use of the following type of combinations 
\beq \label{commut}
 \Tr \left( [\Gamma^{a_1\dots a_i} ,\Deltaphi]K\right) \equiv \Tr \left((\Gamma^{a_1\dots a_i} \Deltaphi -  {\Deltaphi}\Gamma^{a_1\dots a_i} ) K \right)=\Tr \left(\Gamma^{a_1\dots a_i} \Deltaphi K- K {\Deltaphi}\Gamma^{a_1\dots a_i} \right) \ .
\eeq
and similarly for the anticommutator. 

\begin{subsubsection}{${\cal D}K_1'$}

We want to show that susy requires (\ref{N=1EGG2}) and (\ref{vectorDK1}). We recall that as shown in (\ref{nonzerocomp}), $K_1$ has only nonzero components with an odd number of internal indices.

The idea is to reconstruct the twisted derivative of the corresponding $K'$ appearing in each of the equations by summing an equation coming from internal gravitino (which gives a covariant derivative of $K$ with no dilaton or warp factors) together with equations coming from external gravitino plus dilatino, which contribute the required derivatives of dilaton and warp factor.

We start by showing that susy requires (\ref{aDK1}) to vanish. We use the following combination of equations: (\ref{iu1}) coming from internal gravitino,  (\ref{egra1}) and (\ref{egra2}) from external gravitino, and (\ref{dila1}), (\ref{dila2}) from dilatino (the last four multiplied by arbitrary coefficients $n_e$ and $n_d$, that will be set to $n_e=1, n_d=-1$). \begin{align}
0=& -\frac{i}{4}\Tr\left(\Gamma^{mnp2}(e^{A} \Deltap K_1)+\{\Gamma^{mn1}, (n_e\Deltamu +n_d\Deltaphi)\} K'_1\right) \\
=&-2 e^{A-\phi}\nabla_p (e^{\phi} K_1{}^{mnp2})- 2\partial_p(n_eA+n_d\phi)K'_1{}^{mnp2}
\nn \\
&+\frac{1}{2}(1+n_d)H^{mn}{}_p K'_1{}^{1p} +\frac{1}{2}(3+n_d)(*H)^{mn}{}_p K'_1{}^{2}{}_p\nn\\
&-\frac12 e^{-2A+\phi}F_0(4+n_e+5n_d) K'_+{}^{mn12}-\frac14 e^{-2A+\phi}(*F_4)_{pq} (n_e+n_d)K'_+{}^{pqmn} \nn \\
&-\frac12 e^{-2A+\phi}F^{[m|p}(2+n_e+3n_d)K'_{+ \,p}{}^{|n]} \nn \\
=&-2 \nabla_p K'_1{}^{mnp2}+(*H)^{mnp} K'_1{}^{2}{}_p  \nn \\
=&(\mathcal{D} K'_1)^{mn} \nn 
\end{align}
where in the third equality we have used the values $ n_e=1, n_d=-1$.

To show that (\ref{aDK2}) vanishes, we use
\begin{align} \label{dkmd}
0=& -\frac{1}{4}\Tr\left(2\Gamma^{2}{}_{[m} (e^{A}\Delta_{n]}  K_1) -i[\Gamma^{mn1},n_e\Deltamu +n_d\Deltaphi ] K'_1 \right)  \\
=&-2  e^{A-\phi}\nabla_{[m} (e^{\phi} K_1{}^2{}_{n]})-2\partial_{[m}(n_eA+n_d\phi)K'_1{}^{2}{}_{n]} \nn\\
&-H_{pq[m}K'{}_1{}^{1pq}{}_{n]} (1+n_d) +\frac12 e^{-2A+\phi} *F_6 \, (-2+n_e-n_d) K'_+{}_{mn}{}^{12} \nn \\
&+ \frac14 {e^{-2A+ \phi}} F_{pq} (2+n_e+3n_d) K'_+{}^{pq}{}_{mn}+\frac12 {e^{-2A+\phi}} (*F_4)_{[m|}{}^{p}(n_e+n_d)K'_{+\,p|n]} \nn \\
=&-2 \nabla_{[m} K_1'{}^2{}_{n]} \nn \\
=& (\widetilde{{\mathcal{D} K'_1}})_{mn} \nn 
\end{align}
where we have chosen again $n_e=1, n_d=-1$.

To show that (\ref{aDK3}) vanishes, we use
\begin{align}
0=&-\frac{i}{4}\Tr\left(i\Gamma^{n}{}_{1}( e^{A}\Delta_n K_1)+\Gamma^{2}(n_d\Deltaphi+n_e\Deltamu)K_1'\right) \\
=&- e^{A-\phi}\nabla_n(e^{\phi} K_1{}^{n}{}_{1})-\partial_p(n_eA+n_d\phi)K'_1{}^{p}{}_{1}- \frac{1}{6}H_{pqr}(3+n_d)K_1'{}^{2pqr}\nn\\
&+\frac14{e^{-2A+\phi}} \Big[F_{pq}(2+n_e+3n_d)+i(*F_4)_{pq}(n_e+n_d)\Big]K_+'{}^{pq12} \nn \\
=&-\nabla_nK'_1{}^{n}{}_{1}-\frac{1}{3}H_{pqr} K'_1{}^{2pqr} \nn \\
=&(\widetilde{\mathcal{D} K'_1})_{12}
\end{align}
where we have used again $n_e=1, n_d=-1$.\\

For the vectorial equation (\ref{aDK4}), we use (\ref{iu3}) and (\ref{extK3r}) to get
\begin{align}
0&=\Tr\left(-e^A \Delta_m  K_3   + K_0' \Delta_e \Gamma^{m} \right) \\
&=-4 e^{A} \partial_pA K_3^{mp} + e^{\phi} F_{0} K'_1{}^{m}{}_{1}-e^{\phi} (*F_{4})^{mn}K'_1{}^{2}{}_{n}-e^{\phi} F_{np} K'_1{}^{2npm} \nn \\
&=-4 e^{A} \partial_pA K_3^{mp}+ (\mathcal{D} K'_1)^{m1} \nn
\end{align}
where we have used $K_2=K_1 \Gamma^{12}$ and $K_0 =-iK_3\Gamma^{12}$, and  in the last line we have used (\ref{aDK4}).
For the last equation (\ref{aDK6}) we use (\ref{iu0}) and (\ref{extK3l})
\begin{align}
0&=\Tr\left(e^{A} \Delta_m K_0+i K_3'\Delta_e\Gamma^{m} \right) \\
&=4i e^{A-\phi} \nabla_m(e^{\phi} K_3{}^1{}_2) -8e^{A} \partial_pAK_3{}_{m}{}^{p12}-e^{\phi} {*F_{6}} K'_1{}^{2}{}_m-e^{\phi} F_{mn}K'_1{}{}^{n}{}_{1}+e^{\phi} (*F_{4})^{np}K'_{1\,1npm} \nn \\
&=4i e^{A} \partial_m A \, K_3{}^1{}_2 -8e^{A} \partial_pAK_3{}_{m}{}^{p12}+  (\widetilde{\mathcal{D} K'_1})_{m2} \nn
\end{align}
where in the second equality we have used again $K_0 =-iK_3\Gamma^{12}$, and in the third equality we have used 
(\ref{vectorDK}) (which will be shown to hold below). 

\end{subsubsection}

\begin{subsubsection}{${\cal D}K'_+$}

The other set of equations involves
\beq
K_+'=K_3'+iK_2'=e^{3A}(K_3+iK_2)\ .
\eeq 
From (\ref{nonzerocomp}), we see that $K_+$  with an odd number of internal indices is proportional to $iK_2$,
while for an even number of internal indices, $K_+$ is proportional to $ K_3$.

To show the first equation in (\ref{N=1EGG3}), we use (\ref{iu2}), (\ref{extK2l}) and (\ref{dilK2l}) to get
\begin{align} \label{dkmnu}
0=& \frac{1}{4}\Tr\left(\Gamma^{mnp2} (e^{3A}\Deltap K_2)+i \Gamma^{mn1} (n_e \Deltamu K'_2 + n_{d} \Deltaphi K'_2)\right)   \\
=&-2 e^{3A-\phi}\nabla_p (e^{\phi} K_+{}^{mnp2})-2\partial_p(n_e A+n_d\phi)K'_+{}^{mnp2}+2i\partial_{[m}(n_eA+n_d\phi)K_+'^{2}{}_{n]} \nn\\
&+\frac{1}{2}(1+n_d)H_{mnp}K_+'^{1p}+in_dH_{pq[m}K_+'^{1pq}{}_{|n]}+\frac{1}{2}(3+n_d)(*H)^{mnp}K_+'^{2}{}_{p}\nn\\
&+\frac{e^{\phi}}{4} \left(F_0(n_{e}+5n_{d})- i (*F_6) (-4+ n_{e}-n_{d})\right) K_+'^{mn}\nn \\
&+\frac{e^{\phi}}{4}\left(i F^{mn}(4+n_{e}+3n_d)-(*F_4)^{mn}( n_{e}+n_{d})\right)K_+'^{12}\nn\\
&+\frac{e^{\phi}}{8}\left( i (*F_2)^{mn}{}_{pq}(n_{e}+3n_{d})-F^{mn}{}_{pq}(n_{e}+n_{d})\right)K_+'^{pq}\nn\\
&+e^{\phi}\left(F^{[m|}{}_{p}(n_{}+3n_{d})-i(*F_4)^{[m|}{}_p(-2+n_{e}+n_{d})\right)K_+'^{p12|n]} \nn \ .
\end{align}
and
\begin{align} \label{dkmnd}
0=&  \frac{1}{4}\Tr\left(2 i\Gamma_{2[n}( e^{3A}\Delta_{m]} K_2 )- \Gamma_{mn1} (n_{e}  \Deltamu K_2' +n_{d} \Deltaphi K_2')\right) \\
=&-2 e^{3A-\phi} \nabla_{[m} (e^{\phi} K_+{}^2{}_{n]}) -2i\partial_p(n_e A+n_d\phi)K_+'^{mnp2}-2 \partial_{[m}(n_eA+n_d\phi)K_+'^{2}{}_{|n]}\nn\\
&+i\frac{n_d}{2}H_{mnp}K_+'^{1p}-(1+n_d)H_{pq[m}K'^{1pq}{}_{|n]}+i\frac{n_d}{2}(*H)^{mnp}K_+'^{2}{}_{p}\nn\\
&+\frac{e^{\phi}}{4}\left(iF_0(2+n_{e}+5n_{d})+(*F_6)(n_{e}-n_{d})\right)K'_{+\,mn}\nn\\
&-\frac{e^{\phi}}{4}\left(F_{mn}(n_{e}+3n_{d})+i(*F_4)_{mn}(2+n_{e}+n_{d})\right)K_+'^{12}\nn\\
&-\frac{e^{\phi}}{8}\left((*F_2)_{mnpq}(n_{e}+3n_{d}) +iF_{mnpq}(-2+n_{e}+n_{d})\right)K_+'{}^{pq}\nn\\
&+e^{\phi}\left(iF_{[m|p}(n_{e}+3n_{d})+(*F_4)_{[m|p}(n_{e}+n_{d})\right)K_+{}'{}^{p12}{}_{|n]} \nn \ .
\end{align}
Note that in the NS sector $K_+$ reduces to $K_2$, while in the RR sector it is proportional to $K_3$.
We combine these two, choosing $n_e=\tfrac32, n_d =-\tfrac12$, and we get
\begin{align}
0=&\eqref{dkmnu}-i\eqref{dkmnd} \nn \\
=& -2 \nabla_p K'_+{}^{mnp2} +2 i \nabla_{[m} K_+'{}^2{}_{n]} + (*H)_{mnp} K_+'{}^2{}_p - e^{\phi} (*F_{4} - i F_{2})_{mn} K_+'{}^{12} \nn \\
=& \,\,({\mathcal D} K'_+)_{mn} -i (\widetilde{{\mathcal D} K'_+})_{mn} \nn .
\end{align}

For the $12$ components we use 
\begin{align}
0=& \frac{1}{4}\Tr\left(i\Gamma^{n1} (e^{3A}\Deltan K_2)-i\Gamma^{2} (n_e \Deltamu K_2' + n_{d} \Deltaphi K_2')\right)\nn \\
=&i e^{3A-\phi}\nabla_n (e^{\phi} K_+^{n1})+i\partial_n(n_eA+n_d\phi)K_+'^{n1} +\frac{i}{2}(1+\frac{n_d}{3})H_{pqr}K_+'^{2pqr}\nn\\
&+\frac{e^{\phi}}{4} \left(-F_{0}(6+n_{e}+5n_{d})+i *F_{6} (n_{e}-n_{d})\right)K_{+}'^{12}\nn \\
&-\frac{e^{\phi}}{8}\left(i F_{mn}(n_{e}+3n_{d} )-(*F_{4})_{mn}(-2+n_{e}+n_{d})\right)K_{+}'^{mn} \nn \\
=&i \nabla_n K_+'^{n1}+\frac{i}{3} H_{pqr}K_+'^{2pqr}+e^{\phi}(-F_{0}+i *F_{6}) K_+'^{12} \nn \\
=&  (\mathcal{D}  K_+')_{12} - i (\widetilde{\mathcal{D}  K_+'})_{12} \ .
\end{align}
where we have chosen  $n_e=3, n_d=-1$.

We are left with the vectorial components. The last equation in (\ref{N=1EGG3}) is trivial (see \eqref{DK8}). 
To show the $m1$ component, we use 
 \begin{align}
0&=-\frac{1}{4}\Tr\left(\Gamma^{12}(e^{3A}\Delta_m K_3)+in_e \{\Deltamu,\Gamma_m\} K'_3\right) \nn \\
&=e^{3A-\phi}\nabla_m (e^{\phi} K_+^{12})-n_e \partial_m A K_+'^{12} + i\frac{e^{\phi}}{4}(n_e-1)[-F_0K_+'{}_{m1}+(*F_4)_{mp}K_+'^{2p}+F_{pq}K_+'^{2pq}{}_{m}]\nn \\
&=\,(\widetilde{\mathcal{D}K_+'})_{m1}-\partial_m (4A-\phi) K_+'^{12} \nn 
\end{align}
where we have taken $n_e=1$. 
 
 For the $(\widetilde{\mathcal{D} K'})_{m2}$ equation, we first note that supersymmetry
 requires their RR pieces to vanish by itself, namely 
 \begin{equation}
0=\Tr\left(\Delta_m K_3'\right)=e^{\phi} \left((*F_6)(K_+')_{m2}+F_{mp}K_+'^{1p}+(*F_4)^{pq}(K_+')_{1pqm}\right)= \cf_{RR}\big|_{m2} \nn \ ,
 \end{equation}
 while in the $m1$ equation, the RR piece is proportional to a derivative of the warp factor, i.e.
 \begin{align}
0=&\Tr\left(e^{3A} \Delta^m K_0 \right)=4ie^{3A} \nabla_m K_+{}^1{}_2+ e^{\phi} \left(F_0K_+'{}^{m1}-(*F_4)^{mp}K_+'{}^2{}_p-F_{pq}K_+'^{2pqm}\right) \nn \\
=&4i \partial_m A K'_+{}^1{}_2+ \cf_{RR}\big|^{m1} \nn \ .
\end{align}

Then we use
 \begin{align}
0=& \frac{1}{4}\Tr  \left(i\Gamma^{mp12}(e^{3A} \Deltap K_3)+ [\Gamma^m, n_e \Deltamu+n_d \Deltaphi ]  K_3' \right) + \cf_{RR}\big|^{m1}+4i \partial_m AK'_+{}^1{}_2  \nn \\
=&+ 2e^{3A} \nabla_pK_+^{mp12} +2\partial_p(n_eA+n_d\phi) K_+'^{mp12} -\frac{1}{4}(n_d+2)H^{m}{}_{pq}K_+'^{mp12}+i\frac{e^{\phi}}{4}\Big[(*F_6)(5-n_e+n_d)K_{+}'^{m1}\nn\\&+ F_{mp}(3+3n_d+n_e)K_+'^{2p}+(*F_4)_{pq}(-1+n_e+n_d)K_+'^{2pqm}\Big]+ \cf_{RR}\big|^{m1} +4i \partial_m K'_+{}^1{}_2 \nn\\
=&  \,\, (\mathcal{D}K_+')^{m1}-2\partial_p\phi K_+'^{mp12} +4i \partial_m AK'_+{}^1{}_2 \nn
 \end{align}
where in the last equality we have chosen $n_e=3, n_d=-2$. For the $m2$ component, we use
\begin{align}
0=&\frac{1}{4}\Tr  \left(\Gamma^{p}{}_{m}e^{3A}  \Deltap K_3 -  i[\Gamma_{m12},n_e \Deltamu+n_d \Deltaphi ]K_3' \right) + \cf_{RR}\big|_{m2} \nn \\
=& -e^{3A}\nabla_p K_+{}^{p}{}_{m}-\partial_p(n_eA+n_d\phi)K_+'^{p}{}_m- \frac12 H_{mpq}K_+'^{pq12} (2+n_d)  + i\frac{e^{\phi}}{4}\Big [F_0 (5+n_e+5n_d) K_+'{}_{m2} \nn \\
&+ F_{pq} (1+n_e+3n_d)  K_+'^{1pq}{}_m+(*F_{(4)})_{mp} (-3+n_e+n_d) K_+'^{1p}\Big]+ \cf_{RR}\big|_{m2} \nn \\
=& \,\, (\widetilde{\mathcal{D} K_+'})_{m2}-\partial_p(2A-\phi)K_+'^{p}{}_m+H_{mpq}K_+'^{pq12} \nn
\end{align}
where here we have inserted $n_e=5, n_d=-2$.
 
\end{subsubsection}




\end{document}